\documentclass[aps,pre,superscriptaddress,showpacs,preprint]{revtex4-1}

\usepackage{complexity}
\usepackage{graphicx}
\usepackage{amsfonts}

\begin{document}

\title{Making Classical Ground State Spin Computing Fault-Tolerant}
\author{E. J. Crosson}
\affiliation{Department of Physics, University of Washington, Seattle, WA 98195 USA}
\author{D. Bacon}
\affiliation{Department of Computer Science \& Engineering, University of Washington, Seattle, WA 98195 USA}
\affiliation{Department of Physics, University of Washington, Seattle, WA 98195 USA}
\author{K. R. Brown}
\affiliation{Schools of Chemistry and Biochemistry; Computational Science and Engineering; and Physics, Georgia Institute of Technology, Atlanta, GA 30332 USA}
\date{\today}
\begin{abstract}
We examine a model of classical deterministic computing in which the ground state of the classical system is a spatial history of the computation.  This model is relevant to quantum dot cellular automata as well as to recent universal adiabatic quantum computing constructions.  In its most primitive form, systems constructed in this model cannot compute in an error free manner when working at non-zero temperature.  However, by exploiting a mapping between the partition function for this model and probabilistic classical circuits we are able to show that it is possible to make this model effectively error free.  We achieve this by using techniques in fault-tolerant classical computing and the result is that the system can compute effectively error free if the temperature is below a critical temperature.  We further link this model to computational complexity and show that a certain problem concerning finite temperature classical spin systems is complete for the complexity class Merlin-Arthur.  This provides an interesting connection between the physical behavior of certain many-body spin systems and computational complexity.
\end{abstract}
\pacs{05.20.-y,05.10.-a,05.70.Fh,89.70.Eg}
\maketitle

The creation of massive digital and deterministic computing systems represents one of the greatest triumphs of the last century.  Increasingly, however, as the components in our information processing devices shrink to atomic scales~\cite{Feynman:60a}, the two defining characteristics of these systems---that they are digital as opposed to analog, and that they are deterministic as opposed to probabilistic---are beginning to be revealed as approximations that arise from considering macroscopic or mesoscopic physical systems.  As computing elements are made smaller they are increasingly subject to noise and an inability to be perfectly controlled~\cite{Kish:02a}.  Even further down this route lie quantum computers~\cite{Feynman:85a,Benioff:82a,Benioff:82b,Deutsch:85a,Shor:94a} where, instead of probabilistic time evolution, the computer is made of elements which function according to the laws of quantum information.  These trends point toward the necessity of studying computing elements in settings where digital and deterministic functionality are not a priori guaranteed~\cite{vonNeumann:56a}.  Here we consider a physical method for classical computing wherein a computation is encoded into the ground state of a classical many-body system of spins.   At zero temperature in this model, computation is deterministic and thus we label this model classical {\em ground state spin computation}.  The model has many predecessors, include quantum-dot cellular automata~\cite{Lent:93a,Lent:94a,Wang:04a}, the broadcast model on trees~\cite{Evans:00a}, quantum ground state computing~\cite{Mizel:02a,Mizel:04a}, and universal adiabatic quantum computing~\cite{Aharonov:04a,Mizel:06a,Kempe:06a,Oliveira:05a,Love:08a}.  We focus our work here on this model when the system has thermalized at a non-zero temperature.  At non-zero temperature, if we do nothing to compensate for the effects of thermal equilibrium, ground state spin computation fails.  Here we show that it is possible to use the ideas of fault-tolerant classical computing to design ground state spin computers which function deterministically with high probability at non-zero temperatures below a critical temperature.  We achieve this by mapping the thermal ensembles of our computational spin systems onto the distribution of ensembles produced during classical probabilistic computations.  This is the main result of the paper: some classical spin systems in thermal equilibrium can be thought of as enacting classical probabilistic computations spatially across the system.  


Finally, our mapping leads naturally to problems which are complete for the complexity class promise Merlin-Arthur~\cite{Babai:85a,Goldwasser:86a}, a result which can be regarded as a finite temperature version of the Cook-Levin theorem~\cite{Cook:71a,Trakhtenbrot:84a} from the theory of computational complexity.  The new computational model we consider thus serves as a bridge between statistical mechanics and the classical computational complexity of probabilistic computation.  We also obtain a computational complexity result concerning the difficulty of identifying when a system admits a mapping to a probabilistic circuit by showing that a restricted version of this problem is \NP-complete.  This implies that the problem of identifying when a physical system can be seen to be performing a computation is itself computationally intractable.  

The outline of our paper is as follows.  In Section~\ref{sec:cgssc} we introduce the model of computing with spins in the ground state using a particular energy function for this system.  In Section~\ref{sec:1dising} we show that at non-zero temperature the model from the previous section fails to properly compute.  We do this for an extremely simple and previously explored model, but present the result using two different methods, one related to reinterpreting the transfer matrix and the other related to classical circuits applied to simple thermal ensembles.  Generalizing these methods allows us to discuss the energy functions arising in Section~\ref{sec:cgssc} as probabilistic circuits.  This more general formulation we discuss using two different methods in Section~\ref{sec:probtrans}.  Motivated by the mappings discovered in Section~\ref{sec:probtrans}, we then return to our original model and define the broader class of energy functions consistent with these models in Section~\ref{sec:general}.   At this point we also point out how our models differ from classical models of quantum-dot cellular automata. The more general setting leads us to ask questions about how algorithmically hard it is to decide if a given physical system supports computation in these models.  We show that this problem, even in a very weak form, is \NP-complete and thus likely to be intractable.  In Section~\ref{sec:ft} we discuss how to design ground state spin models that, unlike the generic case, do compute fault-tolerantly at non-zero temperature.  Our construction is intimately related to the original fault-tolerance construction of von Neumann~\cite{vonNeumann:56a} and we show that von Neumann's threshold for fault-tolerance is related to a critical temperature in our system.  Finally in Section~\ref{sec:complexity} we discuss how the model we consider is related to the Cook-Levin theorem from computational complexity and show how this leads to certain natural problems about our model being {\bf Promise-}\MA-complete.  

\section{Classical Ground State Spin Computing} \label{sec:cgssc}

Here we introduce the model of classical ground state spin computing.  This model, in particular restricted cases, is relevant to a variety of different models considered elsewhere.  For instance, quantum dot cellular automata can be partially modeled by classical ground state spin computing~\cite{Lent:93a,Lent:94a}.  However the model we consider is more general than these specific instances.  This larger generality comes along with certain unphysical assumptions: our models contain, for example, three-body interactions which are not necessarily easily achievable in a physical device.  Physically we imagine that models such as the one we consider might emerge in certain limits where effective many-body interactions can emerge.  However, irrespective of the physical implementation, the model provides a set of classical many-body interacting systems which can be mapped onto probabilistic classical circuits.  This connects statistical mechanics with classical probabilistic circuits thus bridging computer science and physics in a new and interesting manner.  

We further note that a quantum version similar to this model has been studied by Mizel {\em et al}~\cite{Mizel:02a,Mizel:04a}.  This later model contains significant difference owing to the quantum nature of the computation: in particular the ground state does not contain the computation laid out spatially, but instead exists in a superposition of the computation being carried out spatially.    These quantum models are also connected to universal adiabatic quantum computing schemes~\cite{Aharonov:04a,Mizel:06a,Kempe:06a,Oliveira:05a,Love:08a} and piecewise variations on these schemes~\cite{Bacon:09a,Bacon:09b,Oreshkov:09a,Oreshkov:09b}.   By studying the classical analogy of these schemes, we hope to shed light on how these later quantum methods can be made fault-tolerant.  

\subsection{The Model}

A {\em combinatorial circuit}, $C=(G,L)$, is a directed acyclic graph, $G=(V,E)$ with vertices $V$ and edge set $E$, where the internal vertices, $V_{in}$, of the graph are logical gates, $L:V_{in} \rightarrow {\mathcal G}$ (${\mathcal G}$ is the set of all logical gates), and the external vertices, $V_{ex}$, are the inputs and outputs to the circuit.  External vertices that have no edges leading to them are inputs and external vertices that have no edges that lead away from them our outputs to the circuit.  If a circuit has $n$ input vertices and $m$ output vertices, then to such a circuit we can assign a function $f:\{0,1\}^n \rightarrow \{0,1\}^m$ which results from propagating the input through the logic gates to the output vertices.  Note that at this point we require that fan-outs in our circuit are represented by gates and do not allow circuits which have fan-in gates.  A $k$-fan-out gate is the boolean function $f_k:\{0,1\} \rightarrow \{0,1\}^k$ given by $f_k(x)=(x,\dots,x)$.

Consider the following physical system.  For each edge of a combinatorial circuit associate a single subsystem with only two possible states, $0$ and $1$, i.e. associate a bit to every edge.  Consider next a logic gate which has incoming edges labeled by bits $i_1,i_2,\dots,i_j$ and outgoing edges labeled by $t_1,t_2,\dots,t_k$.  For such a logic gate, $l$, define the following energy function
\begin{equation}
E_l(i_1,i_2,\dots,i_j,t_1,t_2,\dots,t_k)= \left \{ \begin{array}{ll} 0 &{\rm if}~f_l(i_1,i_2,\dots,i_j)=(t_1,t_2,\dots,t_k) \\
\Delta &{\rm otherwise}\end{array} \right., \label{eq:energy}
\end{equation}
where $f_l$ is the function the logic gate is computing.  Note that this energy contributes $0$ when the inputs and outputs follow the rules of the logic gate, but is $\Delta$ otherwise.  Now define the energy for a particular configuration of the bits associated with edges as the sum over all logic gates of these energy terms:
\begin{equation}
E_C(s)=\sum_{l \in L} E_l
\end{equation}
where each $E_l$ acts on the appropriate input and output bits and $s$ is the collection of bits for the entire system.  This energy is a function of the labels on all of the edges.  Its zero energy configurations are ground states which consist of configurations which correctly compute the logical function corresponding to the circuit $C$.  Note that at this point the ground state is degenerate: every valid input (and corresponding output) defines a valid zero energy configuration.  We call an energy function constructed in this fashion from a circuit a {\em circuit energy function}.

Suppose further that, in addition to imposing energy constraints to enforce logic gates, we also impose energy constraints to fix a particular input to the circuit.  Suppose that the input vertices are $v_1,v_2,\dots,v_n$ and we wish to force the input $x_1,x_2,\dots,x_n$, where $x_i \in \{0,1\}$.  Then to each edge whose bit is labeled $s_i$ lead away from vertex $v_i$ we can associate an energy term $E(s_i)=0$ if $s_i=x_i$ and $E(s_i)=\Delta$ otherwise.  Taking a sum over all of these terms for the input vertices (and corresponding edges) we can thus add a term such that the ground state configuration consists now only of the input $x_1,\dots,x_n$ propagated to the output of the circuit, $f(x_1,\dots,x_n)$.  In particular for input $x \in \{0,1\}^n$ define the 
\begin{equation}
E_{C,x}(s) = E_C(s) + \sum_{i=1}^n E_{x_i}(s_i) \equiv  E_C(s) + E_x(s),
\end{equation} 
where 
\begin{equation}
E_{x_i}(s_i) = \left\{ \begin{array}{l}  0~{\rm if}~s_i=x_i~{\rm where}~s_i~{\rm corresponds~to~input}~v_i  \\
\Delta~{\rm if}~s_i \neq x_i~{\rm where}~s_i~{\rm corresponds~to~input}~v_i
\end{array} \right. .
\end{equation}
We call an energy function made up of a circuit energy function plus an input forcing energy function a {\em computed circuit energy function.}

The above description of how to take a combinatorial circuit, $C$, plus its input, $(x_1,\dots,x_n)$, and converting it into a many-spin energy function is what we term the {\em classical ground state spin computing} (CGSSC) model.  Just as in quantum dot cellular automata~\cite{Lent:93a,Lent:94a}, the computation occurs when the system is in its ground state and is spatially spread out across the device.  Unlike in quantum dot cellular automata, however, this model can implement logic gates by using interactions beyond just pairwise interacting spins (or pseudo-spins in the case of the quantum dot configurations.)  For example, constructing a quantum wire in the CGSSC model will correspond directly to an identical model in the semi-classical limit of quantum dot cellular automata models.  Gates in quantum dot cellular automata, however, are implemented in a way which is not directly analogous to the CGSSC model.  In section~\ref{sec:general} we return to this issue and define a more general set of energy functions wherein our results still hold and compare this with quantum cellular automata models.  

\subsection{Example}

\begin{figure}[h]
\begin{center}
\includegraphics{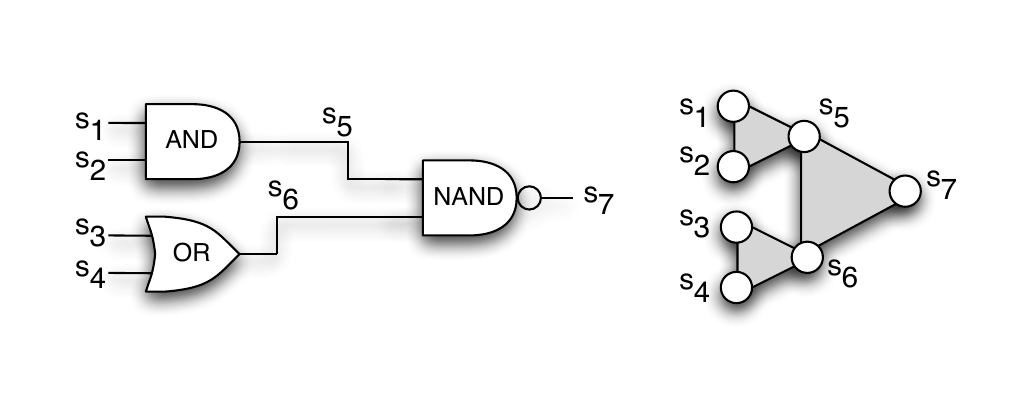}
\caption{Left: An example circuit whose circuit energy function is given in the text.  The bits of the ground state spin energy function are labeled $s_i$.  Right: the spin system corresponding to this circuit.  Here the spins connected by a triangles have interactions between them as specified in the main text.}
\label{fig:circuit}
\end{center}
\end{figure}

To be explicit, let's consider an example circuit involving four inputs, a few logical gates, and one output as diagramed in Fig.~\ref{fig:circuit}.  The energy function for the AND gate is then, for example,
\begin{eqnarray}
E_{AND}(s_1,s_2,s_3,s_4,s_5) &=&\Delta ( (1-s_1)(1-s_2) s_5+(1-s_1)s_2 s_5+s_1(1-s_2) s_5+s_1s_2 (1-s_5)) \nonumber \\
&=&\Delta(s_5+s_1s_2-2s_1s_2s_5).
\end{eqnarray}
The full circuit energy function is given by
\begin{eqnarray}
E_C(s_1,s_2,s_3,s_4,s_5)&=&\overbrace{\Delta(s_3 + s_4 +s_6- s_3 s_4  - 2 s_3 s_6 - 2 s_4 s_6 + 2 s_3 s_4 s_6)}^{\rm OR} \nonumber \\
&&+\underbrace{\Delta (s_5+s_1s_2-2s_1s_2s_5)}_{\rm AND}+\underbrace{\Delta(1-s_7-s_5s_6+2s_5 s_6 s_7)}_{\rm NAND}.
\end{eqnarray}
Suppose that we wish to force the input to be $s_1=s_2=s_3=0$ and $s_4=1$.  Then we would add the term
\begin{equation}
E_{C,x}=E_C+ \Delta(s_1+s_2+s_3+(1-s_4)).
\end{equation}

\section{Unprotected Ground State Spin Computing Fails at Finite Temperature} \label{sec:1dising}

Above we have defined an energy function whose ground state deterministically carries out a circuit.  Having defined this energy model we can now consider the physically important question of what happens to this model when the physical system described by this energy function is in thermal equilibrium at a finite temperature.  Thus we are led to consider the Boltzmann distribution corresponding to a non-zero temperature version of a circuit energy function.  It then makes sense to consider whether the conditional probability of output $f(i_1,i_2,\dots,i_n)=(t_1,t_2,\dots,t_m)$ given the forced input $(i_1,i_2,\dots,i_n)$ for this circuit is large enough to distinguish this output from a completely random outcome.  It is easy to see that for at least some circuits $C$ and inputs $x$, ground state computation will fail to correctly evaluate the function corresponding to the circuit as the size of the circuit being implemented grows.  This follows directly from examining one of the most basic models in statistical physics, the one-dimensional Ising model~\cite{Ising:25a}.  This was pointed out in the context of quantum-dot cellular automata in \cite{Lent:94a} (see also ~\cite{Ungarelli:00a,Wang:04a}).  Here we reproduce this argument presenting it in two different forms.  We do this not just to be pedagogical, but also because these two methods will generalize from the one dimensional case to more general circuit computing energy functions.  

Consider the circuit computing energy function corresponding to inputing the bit $0$ into a series of $n$ identity gates:
\begin{equation}
E_{I}(s_1,\dots,s_{n+1})=\Delta \delta_{s_1,1}+\Delta \sum_{i=1}^n \delta_{s_i, \neg s_{i+1}},
\end{equation}
where $\neg s$ represents the negation of $s$: $\neg 0 =1$ and $\neg 1= 0$.
It is convenient here, and in the sequel, to work with $\{+1,-1\}$ valued variables instead of the $\{0,1\}$ valued $s_i$'s.  Thus we will define uppercase spin variables to be $\pm 1$ valued versions of the $s_i$'s via $S_i=1-2s_i$.  Then the above energy function can be written as
\begin{equation}
E_{I}(S_1,\dots,S_{n+1})= {\Delta \over 2} (1-S_1) + {\Delta \over 2} \sum_{i=1}^n (1-S_i S_{i+1}),
\end{equation}
which we recognize as the one dimensional Ising model with ferromagnetic couplings and a boundary term which is a local field.  The ground state of this is simply all $s_i=0$ ($S_i=+1$), i.e. the initial $0$ has been copied by identity gates down the line.

The thermal ensemble arising from this energy function is given by
\begin{equation}
Pr(S_1,\dots,S_{n+1}) = {\exp(-\beta E_I(S))  \over Z},
\end{equation}
where $Z$ is the partition function,
\begin{equation}
Z=\sum_{S \in \{+1,-1\}^{n+1}} \exp(-\beta E_l(S)),
\end{equation}
and $\beta=(k_B T)^{-1}$ is the inverse temperature. It is well known that the one dimensional Ising model does not order at finite temperature in the thermodynamic limit~\cite{Ising:25a}. From our perspective, we observe that the system will fail to correctly transmit the $0$ down the line at finite temperature except in a window of size $k_B T<{\Delta \over n}$ which goes to zero as the system size goes to infinity.  


\subsection{Transfer Matrix Reinterpreted as a Probabilistic Circuit}

The standard method for solving this model is the transfer matrix method which we review and then reinterpret.  To the energy function add a term dependent on a variable $\gamma$ for the last spin:
\begin{equation}
E_I^\prime(S,\gamma)=E_I(S)+ \gamma S_{n+1}.
\end{equation}  
If we calculate $Z^\prime$ for this energy function, we can then use this to calculate the probability that $S_{n+1}$ is $+1$ via $Pr(S_{n+1}=+1)=\langle \frac{1}{2}(1+S_{n+1}) \rangle= \frac{1}{2} (1+\langle S_{n+1} \rangle)$ where
\begin{equation}
\langle S_{n+1} \rangle =\left. -{1 \over \beta} {\partial \ln Z^\prime \over \partial \gamma} \right|_{\gamma=0}=-\left. {1 \over \beta Z^\prime} {\partial Z^\prime \over \partial \gamma}  \right|_{\gamma=0}.
\end{equation}
Define the two by two matrices
\begin{equation}
[M_i]_{S_{i+1},S_i} = \exp\left[-{b \over 2}(1-S_i S_{i+1} )\right],
\end{equation}
where $b=\beta \Delta$, as well as the column vector
\begin{equation}
[v]_{S_1}=\exp\left[-{b \over 2}(1-S_1) \right],
\end{equation}
and the row vector,
\begin{equation}
[w^T]_{S_{n+1}} = \exp[- \beta \gamma S_{n+1}].
\end{equation}
Explicitly the partition function $Z^\prime$ is
\begin{equation}
Z^\prime=\sum_{S \in \{+1,-1\}^{n+1}} \exp\left[-\beta \gamma S_{n+1}-{b \over 2}\left( 1-S_1+ \sum_{i=1}^n (1-S_i S_{i+1}) \right)\right].
\end{equation}
This can then be written as
\begin{equation}
Z^\prime=\sum_{S \in \{+1,-1\}^{n+1}}  [w^T]_{S_{n+1}} \prod_{i=n}^1 [M_i]_{S_{i+1},S_{i}} v_{S_1} = w^T M_n M_{n-1} \dots M_{1} v,
\end{equation}
where the last equation is a row vector, a matrix product, and a column vector thus giving a scalar.
This is the transfer matrix form of the solution: the partition function is written as a product of ``transfer matrices'' and, in this case, sandwiched between an ``initial'' and ``final'' vector.

Define the following matrices and vectors:
\begin{eqnarray}
[P_i]_{S_{i+1},S_i}&=&{[M_i]_{S_{i+1},S_i} \over 1+ \exp(-b)} \\
\left[p_{in}\right]_{S_{1}}&=& { [v]_{S_1} \over 1+\exp(-b)}.
\end{eqnarray}
We can rewrite $Z^\prime$ as
\begin{equation}
Z^\prime = (1-\exp(-b))^{-(n+1)}w^T P_n P_{n-1} \dots P_1 p_{in}.
\end{equation}
It is at this point that we begin to see our {\em reinterpretation} of the transfer matrix method.  In particular we note that $p_{in}$ is a vector of probabilities (that is it sums to unity) and $P_i$ is a stochastic matrix (its columns sum to unity).  In other words encoded into $Z^\prime$ is a classical preparation of a probabilistic bit, followed by an evolution according to probabilistic gates.  Suppose that we let $p_{out}=P_n P_{n-1} \dots P_1 p_{in}$ denote the output of this probabilistic circuit.  Then 
\begin{equation}
Z^\prime =(1-\exp(-b))^{n-1}\langle w \rangle_{p_{out}}
\end{equation}
where the sample is taken from the output of the probabilistic circuit.  Notice that at $\gamma=0$, $w$ is a vector made up of all components equaling $1$.  In this case, $\left.\langle w \rangle_{p_{out}} \right|_{\gamma=0}= 1$ since then this term is a sum over all the outputs to the probabilistic circuit.  Further
\begin{equation}
\left.-{1 \over \beta}{\partial Z^\prime  \over \partial \gamma}  \right|_{\gamma=0} =(1-\exp(-b))^{n+1}\langle \sigma \rangle_{p_{out}}
\end{equation}
where $\sigma=(+1,-1)^T$.  Thus
\begin{equation}
\langle S_{n+1} \rangle = \langle \sigma \rangle_{p_{out}}.
\end{equation}
In other words the expectation value of the last spin in the chain is equal to the expectation value for the output of the circuit starting with $p_{in}$ and then applying the probabilistic gates $P_1,\dots,P_{n}$.  Note that the combinatorial factors $(1+\exp(-b))^{n+1}$ have canceled out.  

The above description tells us that we can think about the one dimensional Ising spin chain with a forced boundary term as a probabilistic circuit related to our ground state spin computation.  In particular the probabilistic circuit starts with the possibility of an incorrectly initialized boundary.  This is formulated in the preparation probability vector $p_{in}$.  Then with probability $\exp(-b) \over 1-\exp(-b)$ the ground state spin computation of an identity gate is replaced by a bit flip gate.  The probability of the final spin of the chain being in a particular state is then directly given by the probability that the probabilistic circuit outputs a particular value.   

In this model the probability of the ground state properly computing the desired identity circuit is such that the information is washed out at finite temperature.  To see this we must actually calculate $\langle S_{n+1}\rangle$.  This can be done most easily by diagonalizing the gates.  In particular if we define the Hadmard matrix,
\begin{equation}
H={1 \over \sqrt{2}} \left[ \begin{array}{cc} 1 & 1 \\ 1 & -1 \end{array} \right]
\end{equation}
then
\begin{equation}
P_i=H D H^{-1} = H D H
\end{equation}
where
\begin{equation}
D= \left[ \begin{array}{cc} 1 & 0 \\ 0 & {1-\exp(-b) \over 1+ \exp(-b)}\end{array} \right].
\end{equation}
Thus
\begin{equation}
\langle S_{n+1} \rangle = \sigma^T H D^{n} H p_{in} 
\end{equation}
which can be reduced to
\begin{equation}
\langle S_{n+1} \rangle = \left( {1 - e^{-b} \over 1+e^{-b}} \right)^{n+1}=\left[\tanh {b \over 2} \right]^{n+1}.
\end{equation}
From this expression we see that the probability that the final spin is aligned with the input quickly drops to $1/2$ as a function of $n$ (unless $b>n$, i.e. in a small window of temperature below $\frac{\Delta}{n}$.)  Thus the circuit fails with high probability as we make the chain longer and longer.  This is exactly the result of~\cite{Wang:04a}.  In this respect, ground state spin computing in this model is not tolerant to errors arising from working at finite temperature.  Note that we do not consider scaling of the temperature to stay within the small window where the computation is correctly performed to be a fault-tolerant method as it requires unreasonable physical resources as the system size gets larger. 

\subsection{A Second Approach Using Controlled-Not Gates} \label{sec:1dgate}

Having shown in the previous subsection how one can reinterpret the transfer matrix method as a probabilistic circuit for the one dimensional Ising model with forced boundary, we next present a second way to derive this observation.  This method is a direct variation on the method used in~\cite{Eggarter:74a} to study Ising models on Cayley trees.
 
On the $n+1$ spins, define the controlled-not on positions $i$ and $i+1$ as the function, $C_i$, which maps $\{+1,1\}^{n+1}$ to $\{+1,-1\}^{n+1}$ via
\begin{equation}
C_i(S_1,\dots,S_i,S_{i+1},S_{i+2}\dots,S_n)=(S_1,\dots,S_i, S_{i+1} S_i,S_{i+2},\dots,S_n).
\end{equation}
We call this a controlled-not because it functions as a deterministic gate which, controlled on the spin $S_i$ either does nothing to the spin in position $(i+1)$ (if $S_i=+1$), or flips the spin in position $(i+1)$ (if $S_i=-1$.)  Further define the function which is the composition of controlled-not's starting at the first spin and working forward (note that $C_i \circ  C_{i+1} \neq C_{i+1} \circ C_i$):
\begin{equation}
C=C_{n} \circ C_{n-1} \circ \cdots \circ C_2 \circ C_1.
\end{equation}
Notice that if we start with spin $1$ in the state $S_1$ and all of the other spins $S_2,\dots,S_{n+1}$ in $+1$, then the action of $C$ is to {\em copy} $S_1$ down the line: $C(S_1,+1,\dots,+1)=(S_1,S_1,\dots,S_1)$.  In this sense $C$ is the operation of applying identity gates as information is propagated down the spin chain.  Note that the $C_i$ are bijections as is $C$ and that $C_i$ is its own inverse.  Thus
\begin{equation}
C^{-1}=C_1 \circ C_2 \circ \dots \circ C_{n-1} \circ C_{n}.
\end{equation}
Explicitly,
\begin{equation}
C(S_1,S_2,S_3,\dots,S_{n+1}) =(S_1,S_1S_2,S_1S_2S_3, \dots, \prod_{i=1}^{n+1} S_i)
\end{equation}
and
\begin{equation}
C^{-1}(S_1,S_2,S_3,\dots,S_{n+1})=(S_1,S_1S_2,S_2S_3,\dots,S_n S_{n+1}). \label{eq:cnotinverse}
\end{equation}

Consider the energy function on $n+1$ spins:
\begin{equation}
E^\prime(S_1^\prime,S_2^\prime,\dots,S_{n+1}^\prime) = {\Delta \over 2} \sum_{i=1}^{n+1} (1-S_i^\prime).
\end{equation}
Then clearly a system in the thermal state described by this energy function has each spin in $+1$ with probability $1-p=1/(1+\exp(-b))$ and $-1$ with probability $p=\exp(-b)/(1+\exp(-b))$.   Suppose that we apply the $C$ function to this system.  That is consider the above system in its thermal ensemble and consider physically applying the controlled-not gates.  Then certainly this can be thought of as a system in which the first bit is prepared in $0$ with probability $1-p$, with probability $1-p$ the controled-nots successfully copy the bit down the line, and with probability $p$ the output of the controlled-not is flipped.  This is exactly the probabilistic circuit as described in the previous section. 

How is this related to our original $E$?  Let the unprimed variables be the spins after $C$ has been applied to the system.  Then
\begin{equation}
Pr(S_1,\dots,S_{n+1}) = \sum_{S_1^\prime,\dots,S_{n+1}^\prime \in \{\pm 1\}} Pr(S_1,\dots,S_{n+1}|S_1^\prime, \dots, S_n^\prime) Pr(S_1^\prime,\dots,S_{n+1}^\prime)
\end{equation}
where
\begin{equation}
Pr(S_1,\dots,S_{n+1}|S_1^\prime, \dots, S_n^\prime) = \delta_{C(S_1^\prime,\dots,S_{n+1}^\prime),(S_1,\dots,S_{n+1})}= \delta_{(S_1^\prime,\dots,S_{n+1}^\prime),C^{-1}(S_1,\dots,S_{n+1})}.
\end{equation}
Therefore
\begin{equation}
Pr(S_1,\dots,S_{n+1}) = Pr(C^{-1}(S_1,\dots,S_{n+1})).
\end{equation}
This implies that the probability distribution produced by taking the thermal ensemble for $E^\prime(S_1^\prime,\dots,S_{n+1}^\prime)$ and applying $C$ to the system is equivalent to a thermal ensemble with a new energy function
\begin{equation}
E^\prime(C^{-1}(S_1,\dots,S_{n+1})).
\end{equation}
A quick calculation using Eq.~(\ref{eq:cnotinverse}) finds that
\begin{equation}
E^\prime(C^{-1}(S_1,\dots,S_{n+1}))={\Delta \over 2} (1-S_1) + {\Delta \over 2} \sum_{i=1}^n (1-S_i S_{i+1})
\end{equation}
which we see is equal to our original energy function $E$.  Reversing the above argument we thus see that the energy function $E$ can equally well be thought of as the ensemble corresponding to $E^\prime$ followed by the application of the mapping $C$.  Further this later ensemble and computation has a clear interpretation in terms of a probabilistic initialization followed by probabilistic gates: when we apply controlled-nots with error target values then this causes a bit flip error in copying the information down the line.  Thus we see that we can derive the same probabilistic gate expression for the one-dimensional Ising model as that which arose by reinterpreting the transfer matrix as a probabilistic circuit.

\section{The Partition Function as a Probabilistic Circuit}\label{sec:probtrans}

We now turn to the more general setting of a circuit energy function for a generic circuit $C=(G,L)$.  We begin by focusing on circuit energy functions without a forced input.

Let $E_C$ be the energy function for the circuit $C$.  For each logic gate, $l$, with inputs $i_1,i_2,\dots,i_j$ and output $t_1,t_2,\dots,t_k$, define the following {\em tensor},
\begin{equation}
M^{(l)}_{i_1,\dots,i_j,t_1,\dots,t_k}= \left \{ \begin{array}{ll} 1&{\rm if}~f_l(i_1,i_2,\dots,i_j)=(t_1,t_2,\dots,t_k) \\
e^{-b} &{\rm otherwise}\end{array} \right., \label{eq:mtensor}
\end{equation}
where $b=\beta \Delta$ as before. Given this definition, we can now write the partition function $Z$ as the contraction of a  tensor network~\cite{Markov:08a} plus a sum over all inputs and outputs.  What is a tensor network?  Take a graph and map a tensor to every vertex in such a manner that the tensor has an index for every edge of the relevant vertex.  One can then perform a sum over two tensor indices connected via edges in the graph.  A tensor network is thus a graph where vertices are tensors, and edges are tensor indices.  Free edges are indices which have not been summed over and connected edges are ones for which a sum has been performed.  The entire network (graph) then represents itself a tensor with a number of unsummed indices equal to the number of free edges. 

Given the tensors $M^{(l)}$ and the graph given by the circuit (which will have free edges for the inputs and outputs), we can build a tensor network out of these $M^{(l)}$s.  This tensor network will itself be a tensor network having indices for the inputs and outputs.  Call this tensor $M_{i_1,\dots,i_n,t_1,\dots,t_m}^C$ for inputs $i_1,\dots,i_n$ and output $t_1,\dots,t_m$.  Then it is easy to see that the partition function is equal to
\begin{equation}
Z=\sum_{i_1,\dots,i_n,t_1,\dots,t_m \in \{+1,-1\}} M_{i_1,\dots,i_n,t_1,\dots,t_m}^C
\end{equation}
In other words, the partition function is given by the sum over all inputs, and all outputs of the value of the tensor network.

What does this have to do with probabilistic circuits?  Well now instead of using the tensors $M$, instead use
\begin{equation}
P^{(l)}_{i_1,\dots,i_j,t_1,\dots,t_k}={ M^{(l)}_{i_1,\dots,i_j,t_1,\dots,t_k} \over 1+(2^k-1)e^{-b}} \label{eq:ptensor}
\end{equation}
or in other words
\begin{equation}
P^{(l)}_{i_1,\dots,i_j,t_1,\dots,t_k}=\left \{ \begin{array}{l} {1\over 1+(2^k-1)e^{-b} }~{\rm if}~f_l(i_1,i_2,\dots,i_j)=(t_1,t_2,\dots,t_k) \\
{e^{-b}\over 1+(2^k-1)e^{-b} }~{\rm otherwise}\end{array} \right.
\end{equation}
It is easy to check that this defines a {\em stochastic} matrix, i.e. a probabilistic gate.  Indeed it is a probabilistic gate which performs the correct function of the gate with probability ${1 \over 1+(2^k-1)e^{-b} }$ and randomly flips the output to one of the other outputs with equal probability ${e^{-b}\over 1+(2^k-1)e^{-b} }$.  

Notice, importantly, that the difference between $P^{(l)}$ and $M^{(l)}$ is a constant which depends on only on the number of output bits.  So suppose we consider the tensor network for the circuit made now with $P^{(l)}$s and not $M^{(l)}$s.  Then we can write the partition function as
\begin{equation}
Z_C=K \sum_{i_1,\dot,i_n,t_1,\dots,t_m \in \{+1,-1\}} P_{i_1,\dots,i_n,t_1,\dots,t_m}
\end{equation}
 where $K$ is the simple combinatorial factor
\begin{equation}
K=\prod_l (1+(2^{k_l}-1)e^{-b})
\end{equation}
and $k_l$ is the number of outputs of the $l$th gate.  Now examine $P_{i_1,\dots,i_n,t_1,\dots,t_m}$.  This is nothing more than the probability that we get output $t_1,\dots,t_m$ given that we had input $i_1,\dots,i_n$ to the probabilistic circuit with probabilistic gates $P^{(l)}$.  In other words, we can express the partition function as
\begin{equation}
Z_C=K \sum_{inputs} \sum_{output} {\rm Pr(output|input)}
\end{equation}
Thus the partition function is really hiding a probabilistic computation.  Notice further that we can explicitly calculate that partition function because $\sum_{output} {\rm Pr(output|input)}=1$,
\begin{equation}
Z_C= K 2^n
\end{equation}
where $n$ is the number of input bits.

Now we turn to the case where we force the input to the circuit. Let $x_1,\dots,x_n$ be the inputs to the circuit $C$ and $E_{C,x}$ be the energy function for the forced computation.  Define the modified energy function
\begin{equation}
E_{C,x,y} = E_{C,x}(s)+E_y(s),
\end{equation}
where
\begin{equation}
E_y(s)= \sum_{i=1}^m \gamma_i E_{y_i}(s)
\end{equation}
with
\begin{equation}
E_{y_i}(s)= \left\{ \begin{array}{ll}\gamma_i   t_i &{\rm if}~y_i=1 \\ \gamma_i  (1-t_i) & {\rm if}~y_i=0 \end{array}\right.,
\end{equation}
and the sum is over the output bits, $t_i$.  Call the partition function associated with this setup $Z_{C,x,y}$.  

Note that $Z_{C,x,y}|_{\gamma_1=\cdots=\gamma_m=0}$ is equal to the partition function for energy function with a forced input $x$, $Z_{C,x}$.  Define
\begin{equation}
P_{in}(i_1,\dots,i_n)= \prod_{j=1}^n P_{in,j}(i_j),
\end{equation}
where
\begin{equation}
Pr_{in,j}(i_j)=\left\{ \begin{array}{ll} {1 \over 1+e^{-b}}  & {\rm if}~i_j=x_j \\
{e^{-b} \over 1+e^{-b}} & {\rm if}~i_j \neq x_j\end{array} \right. .
\end{equation}
Then it is easy to see that
\begin{equation}
Z_{C,x}=K^\prime \sum_{i_1,\dot,i_n,t_1,\dots,t_m \in \{+1,-1\}} P_{i_1,\dots,i_n,t_1,\dots,t_m} P_{in}(i_1,\dots,i_n),
\end{equation}
where
\begin{equation}
K^\prime= K (1-e^{-b})^n.
\end{equation}
Thus we see that we can interpret $Z_{C,x}$ as a sum over a probabilistic circuit, but now with a probabilistic input corresponding to each bit being flipped with a probability $e^{-b} \over 1+e^{-b}$.  Again, because we are summing over all outputs, we can explicitly perform this sum
\begin{equation}
Z_{C,x}=K^\prime.
\end{equation}

We are interested in computing the probability that the thermal ensemble for the forced input circuit has output given by $y_1,\dots,y_n$.  This can be calculated by 
\begin{eqnarray}
Pr(t_1=y_1,\dots,t_m=y_m)&=&\left. -{1 \over \beta Z_{C,x,y}} {\partial^{m} Z_{C,x,y} \over \partial \gamma_1 \cdots \partial \gamma_m} \right|_{\gamma_1=\cdots=\gamma_m=0} \nonumber \\
&=&\left. -{1 \over \beta Z_{C,x}} {\partial^{m} Z_{C,x,y} \over \partial \gamma_1 \cdots \partial \gamma_m} \right|_{\gamma_1=\cdots=\gamma_m=0} 
\end{eqnarray}
Note that
\begin{equation}
Z_{C,x,y}=K^\prime \sum_{i_1,\dot,i_n,t_1,\dots,t_m \in \{+1,-1\}} e^{-\beta E_y(s)}P_{i_1,\dots,i_n,t_1,\dots,t_m} P_{in}(i_1,\dots,i_n)
\end{equation}
so that the partial derivatives may be evaluated only over the first term.  There we find that
\begin{equation}
\left.{\partial^m  e^{-\beta E_y(s)} \over \partial \gamma_1 \cdots \partial \gamma_2} \right|_{\gamma_1=\cdots=\gamma_n=0} =-\beta \prod_{i=1}^m \delta_{t_i,y_i}.
\end{equation}
Thus we find that
\begin{equation}
Pr(t_1=y_1,\dots,t_m=y_m)=\sum_{s_1,\dots,s_n} P_{s_1,\dots,s_n, y_1,\dots,y_n} P_{in}(s_1,\dots,s_n).
\end{equation}
This is the main result of this section: the probability of the ground state spin computing model output bits is equal to the probability of running the probabilistic circuit described by $P$ on inputs described by input probability distribution $P_{in}$.  The gates in $P$ are simply noisy versions of the deterministic gates which fail with a fixed probability related to the temperature.

This result is interesting in two manners.  First it allows one to use techniques designed for probabilistically failing gates to be ported over to our model.  For instance this will allow us to use methods for constructing fault-tolerant circuits.  It is also interesting in that it shows that certain many-body quantum systems can be efficiently simulated.  Computed circuit energy functions can be simulated by directly implementing the probabilistic gate that these systems correspond to.  Here we have shown that output bits for the circuit energy function are related to the output bits for the probabilistic computation.  A straightforward generalization shows that this is also true of other bits in the system.  We can efficiently simulate systems with computed circuit energy functions by executing the probabilistic computation these energy functions represent.

\section{The Gate Model Derivation}

Next let us turn to the same analysis as the prior subsection, but now using the gate trick as we did for the one-dimensional model in Section~\ref{sec:1dgate}.  To do this we must first define the equivalent of the controlled-not gate.  Consider a logical gate $l$ with input $i_1,\dots,i_j$ and output $t_1,\dots,t_k$ which computes the function $(t_1,\dots,t_k)=f(i_1,\dots,i_j)$.  Define the function $C_l$ from $\{0,1\}^j \times \{0,1\}^k$ to $\{0,1\}^j \times \{0,1\}^k$ as
\begin{equation}
C_l(i_1,\dots,i_j,t_1,\dots,t_k)=\left\{ \begin{array}{ll} 
(i_1,\dots,i_j,f_l(i_1,\dots,i_j)) & {\rm if}~t_1=\cdots=t_k=0 \\ 
(i_1,\dots,i_j,0,\dots,0) & {\rm if}~(t_1,\dots,t_k)=f(i_1,\dots,i_j) \\
(i_1,\dots,i_j,t_1,\dots,t_k) &{\rm otherwise}
\end{array}\right.
\end{equation}
This function thus computes the function on input $(i_1,\dots,i_j,0,\dots,0)$, uncomputes the function on input $(i_1,\dots,i_j,f_l(i_1,\dots,i_j))$, and does nothing otherwise.  Note that $C_l$ is a bijection and is self-inverse, $C_l^{-1}=C_l$.  Defining an order to gates in our circuit such that the circuit computes properly, $l_1$ first, $l_2$ second, etc.  Then we can define the function on all our bits of
\begin{equation}
C_C= C_{l_r} \circ \cdots \circ C_{l_2} \circ C_{l_1}
\end{equation}
Suppose we initially start with input bits initialized to an input $i_1=x_1,\dots,i_n=x_n$ and all other bits initialized to $0$.  Then if we apply $C_C$ to these bits we will place the result of the computation in their appropriate locations across the circuit.

Next define an energy function on all of our system.  This is most easily defined in terms of the inputs to the full circuit and the outputs to the gates $l \in L$,
\begin{equation}
E^\prime(s^\prime)= \sum_{i_1^\prime,\dots,i_m^\prime} \Delta(\delta_{i_1^\prime,\neg x_1}+\cdots+\delta_{i_m^\prime, \neg x_m})+\sum_{l \in L} E_l^\prime(t_1^\prime,\dots,t_{k_l}^\prime).
\end{equation}
and
\begin{equation}
E_l^\prime(t_1^\prime,\dots,t_{k_l}^\prime)= \left\{ \begin{array}{ll}
0 &{\rm if}~ t_1^\prime=\cdots=t_{k_l}^\prime=0 \\
\Delta & {\rm otherwise}
 \end{array}\right. .
\end{equation}
This energy function thus assigns for non-global inputs an energy penalty for output to gates being anything different that $0^{k_l}$.  For global inputs it adds a penalty for every input which is not the correct input from $x$.  The thermal ensemble resulting from this system is then trivially described: the input qubits are in the correct input $x_i$ with probability $1 \over 1+\exp(-b)$ and the internal output bits of a gate are grouped together and have a probability $1 \over 1+(2^{k_l}-1)\exp(-b)$ being all zeros and probability $e^{-b} \over 1+(2^{k_l}-1)\exp(-b)$ to be anything else.  

Imagine applying $C_C$ to the state described by the ensemble for $E^\prime(s^\prime)$.  Clearly the ground state of $E^\prime(s^\prime)$ is such that the circuit $C$ will be correctly computed.  Further, because the ensemble related to $E^\prime(s^\prime)$ has erred inputs, the resulting probability distribution will have erred inputs.  Finally the gates will function properly only when the output was properly initialized to all $0$s.  This occurs with probability $1 \over 1+(2^{k_l}-1)\exp(-b)$ for logic gate $l$ and otherwise the gate fails by an equally probable error state with probability $e^{-b} \over 1+(2^{k_l}-1)\exp(-b)$.  In other words the ensemble is exactly the one describing the probabilistic circuit in Section~\ref{sec:probtrans}.  Let us now show that this corresponds to the thermal ensemble of our original $E_C$ construction.  

As in Section~\ref{sec:1dgate} the trick here is to express the probability distribution for the ensemble after applying $C_C$ as
\begin{equation}
Pr(s)=\sum_{s^\prime}Pr(s|s^\prime) Pr(s^\prime),
\end{equation}
where
\begin{equation}
Pr(s|s^\prime) = \delta_{C_C(s^\prime),s}= \delta_{s^\prime,C_C^{-1}(s)},
\end{equation}
such that
\begin{equation}
Pr(s)=Pr(C_C^{-1}(s)).
\end{equation}
Thus the probability distribution resulting from after the application of $C_C$ to the $E^\prime(s^\prime)$ thermal ensemble is equal to the probability distribution arising from $E^\prime(C_C^{-1}(s))$.

We will now show the $E^\prime(C^{-1}(s))$ is exactly the energy function we would define for a ground state spin computing construction for our circuit $C$.  Recall that
\begin{equation}
C_C^{-1}= C_{l_1} \circ C_{l_2} \cdots \circ C_{l_r}.
\end{equation}
Note that in evaluating $E^\prime(C^{-1}(s))$, which is a sum of terms, we can evaluate the action action of $C_C^{-1}$ on each of these terms separately and then re-sum.  Thus we first examine
\begin{equation}
C_C^{-1}\left[ \sum_{i_1^\prime,\dots,i_m^\prime} \Delta(\delta_{i_1^\prime,\neg x_1}+\cdots+\delta_{i_m^\prime, \neg x_m})\right] = \sum_{i_1^\prime,\dots,i_m^\prime} \Delta(\delta_{i_1^\prime,\neg x_1}+\cdots+\delta_{i_m^\prime, \neg x_m})
\end{equation}
This is because the inputs to the circuit commute through the individual gate elements $C_{l_i}$.  Indeed this observation along with the order of $C_C^{-1}$ allows one to derive the term arising from each $E_l^\prime(C_l(i_1,\dots,i_{j_l},t_1,\dots,t_{k_l}))$ term independently.  In particular the only term which is not $\Delta$ for this energy function is 
\begin{equation}
E_l^\prime(C_l(i_1,\dots,i_{j_l},f_l(i_1,\dots,i_{j_l})))=0
\end{equation}
But this simply means that
\begin{equation}
E_l^\prime(i_1,\dots,i_{j_l},t_1,\dots,t_{k_l}) = 0
\end{equation}
only when $(t_1,\dots,t_{k_l})=f_l(i_1,\dots,i_{j_l})$ and is $\Delta$ otherwise.  This is just our normal definition of the circuit energy function.  

Thus we see that using the $C_C$ construction one can construct a probabilistic circuit from the thermal ensemble for $E_{C,x}^\prime(s^\prime)$ which is the probabilistic circuit described in Section~\ref{sec:probtrans}.  Further we have shown that this can be thought of as arising from the original energy function $E_{C,x}$.  This provides an alternative derivation of our main result.

\section{Generalized Classical Ground State Spin Computation Models} \label{sec:general}

Having shown that classical ground state spin computing at non-zero temperature can be mapped to probabilistic circuits it is interesting to return to our original model and consider what generalizations of the energy function allow for this mapping to occur.  The crucial assumption during our derivation of the probabilistic re-interpretation of the partition function was that we could convert the tensors $M^{(l)}$ of Eq.~(\ref{eq:mtensor}) to tensors $P^{(l)}$ of Eq.~(\ref{eq:ptensor}) in such a way that $P^{(l)}$ could be interpreted as a probablistic gate.  This requires (1) the division of tensor indices into input and output indices and (2) for each input to the circuit the normalization to make the outputs probabilities (positive and sum to unity) is the same.  

Let us begin by showing that not all tensors and divisions into input and output vertices can result in a probabilistic interpretation.  This can be demonstrated with a simple example on two bits:
\begin{equation}
E(s_1,s_2)=\left \{ \begin{array}{ll} 0 & {\rm if}~ s_1=s_2=0 \\
\Delta & {\rm if}~s_1=s_2=1 \\
2\Delta & {\rm otherwise} \end{array}\right. .
\end{equation}
If we chose $s_1$ as an input and $s_2$ as an output, this gives rise to the tensor
\begin{equation}
M=\left[ \begin{array}{cc} 
1 & e^{-2 b} \\
e^{-2b} & e^{-b} 
\end{array}\right].
\end{equation}
In order that the first column of this matrix sum to unity we must divide by $1+e^{-2b}$.  However when we do this the second column will not sum to $1$.  A similar argument occurs if we had chosen $s_1$ as the output and $s_2$ as the input.  Thus there is no way to scale $M$ in such a way that it can be interpreted as a probabilistic gate.  

A more relevant example of a failure for a model to not be amenable to our scaling arising in quantum-dot cellular automata models.  In quantum dot cellular automata models bistable quantum dots are engineered in such a way that the ground state enacts a classical computation as in our model.  Most of the treatments of these models deal with the quantum mechanics of these devices~\cite{Lent:93a,Lent:94a,Wu:98a}.  One can, however, build a classical model of these systems as was done, for example, by Wang in \cite{Wang:04a}.  In this approximation one derives a classical statistical mechanical many spin system as in our model.  When one does this for a quantum-dot cellular automata wire one ends up with exactly the model we can consider above in Section~\ref{sec:1dising}.  However when one considers this for the majority gate constructions (see ~\cite{Lent:93a,Lent:94a,Wang:04a}) one obtains a model for which our technique cannot be applied.  In particular the majority gate that they propose has a classical energy function which is 
\begin{equation}
E_M(S_1,S_2,S_3,S_m) = -\Delta(S_1 S_m +S_2 S_m +S_3 S_m)=-\Delta(S_1+S_2+S_3) S_m.
\end{equation}  
Here $S_1,S_2,S_3$ are the input spins and $S_m$ is the output spin.  (Note that we have used one less spin than in the traditional presentation, but this does not affect our conclusions as one can think of the extra spin as being connected to the central spin by an identity gate.)  Note that the lowest energy configuration {\em for a given} input correctly computes the majority function.  That is if we fix $S_1$, $S_2$, and $S_3$, then the lowest energy value of $S_m$ is given by the majority of the inputs, $S_m=MAJ(S_1,S_2,S_3)$.  Note, however, that if we sum over all outputs for input $S_1=S_2=S_3=+1$
\begin{equation}
\sum_{S_m|S_1=S_2=S_3=+1} \exp(-bE_M(S_1,S_2,S_3,S_m))=e^{-3b}+e^{-3b},
\end{equation}
but if one uses input $S_1=S_2=+1$ and $S_3=-1$ one obtains
\begin{equation}
\sum_{S_m|S_1=S_2=+1,S_3=-1} \exp(-bE_M(S_1,S_2,S_3,S_m))=e^{b}+e^{-b}.
\end{equation}
Thus we cannot properly globally normalize the entire tensor in such a way as to to obtain a probabilistic gate.  

However, the same two-spin energy function can be used to describe a fanout gate, in contrast to the many-spin interaction used in Eq.~(\ref{eq:energy}),
\begin{equation}
E_F(S_1,S_2,\dots,S_{k+1})=-\Delta S_1 \sum_{i=2}^{k+1} S_i. \label{eq:fan-out}
\end{equation}
Here $S_1$ is the input and the other spins are the output. By changing the direction of the circuit, we can show that this element can be represented as a probabilistic gate.  I.e. the majority gate used in quantum dot cellular automata can not be mapped to a probabilistic circuit, but reversing the role of the inputs and outputs, this gate can be used to construct a fan-out gate.  To see this note that, as for our previous fanout model, the ground state of this system, where the computation is performed, is degenerate.  However now, as opposed to a single excited state there are multiple excited states.  For example if we consider a fan-out to two spins we obtain an $M$ as given in Table~\ref{tab:energy}.
\begin{table}[h]
\begin{center}
\begin{tabular}{|c|cc|c|}
\hline $S_1$ & $S_2$ & $S_3$  & $e^{-\beta E_F}$  \\
\hline
$+1$ & $+1$ & $+1$ & $e^{2b}$ \\
$+1$ & $+1$ & $-1$ & $1$ \\
$+1$ & $-1$ & $1$ & $1$ \\
$+1$ & $-1$ & $-1$ & $e^{-2b}$ \\
$-1$ & $+1$ & $+1$ & $e^{-2b}$ \\
$-1$ & $+1$ & $-1$ & $1$ \\
$-1$ & $-1$ & $1$ & $1$ \\
$-1$ & $-1$ & $-1$ & $e^{2b}$ \\ \hline
\end{tabular}
\end{center}
\caption{The 2 output fan-out Boltzman factor.}
\label{tab:energy}
\end{table}
From this table we see that while the outputs that are in error have different Boltzman factors, we see that the sum over the output spins is indeed independent of the output.  Indeed when we interpret this as a probabilistic gate by dividing out by $K=2+e^{2b}+e^{-2b}$, then we see that this is a probabilistic gate which, instead of having a global failure where all possible output failures have equal probability, the fan-out spins have independent error probabilities.  In other words it is as if the fan-out occurred and then with probability $e^{-b} \over e^{-b}+e^b$ each of the outputs is {\em independently} flipped.  We will use this fan-out gate in the next section when we discuss fault-tolerance.

What is a necessary and sufficient condition for an energy function to allow for a proper normalization to a probabilistic gate?  If we insist that this work for all temperatures, it is easy to give this answer.  Let $E^{(l)}(i_1,\dots,i_j,t_1,\dots,t_k)$ denote the energy function for a logic gate with inputs $i_1,\dots, i_j$ and output $t_1,\dots,t_k$.  Let ${\mathcal E} (i_1,\dots,i_k) = \{ E(i_1,\dots,i_k,t_1,\dots,t_k)| t_1,\dots,t_k \in \{0,1\}\}$ denote the set of energies for a given input.  Then the necessary and sufficient condition is that the ${\mathcal E} (i_1,\dots,i_k)$'s must be permutations of each other. The sufficiency of this condition is obvious.  Necessity follows from noting that we are requiring this to hold for all temperatures and thus the sum over outputs of the $\exp(-\beta E)$ must be a rearrangement of this sum and therefore a permutation of the output energies.  Note that this condition implies that the ground states of a logic gate energy function must be degenerate for all possible inputs (but note that this alone is not sufficient for the condition to hold.)

\subsection{Computational Complexity of Identifying Probabilistic Circuits in Thermal Systems}

Having established a necessary and sufficient condition for the possibility of scaling the $M^{(l)}$ tensors as probabilistic gates, a natural follow up question is how one can determine whether it is possible to take an energy function $E(s_1,\dots,s_n)$ and determine whether or not it can be interpreted as a probabilistic circuit.  Here we will show that this problem is computationally intractable, even when we are given information about the breakdown of $E(s_1,\dots,s_n)$ into terms which we wish to interpret as gates. 

Suppose that we are told the $E(s_1,\dots,s_n)$ is made up of a sum over logic gates and preparations 
\begin{equation}
E(s_1,\dots,s_n)=\sum_l E_l, 
\end{equation}
i.e. we are given a specification of the $E_l$ in terms of a constant number of bits and we are told that each of these $E_l$ will correspond to a logic gate or a preparation of initial inputs.  We can then ask: can we interpret this as a probabilistic circuit with or without prepared inputs?  Every $E_l$ involves some inputs and output bits, but in general we will not be told which of these are inputs and which are outputs.  In general an $M^{(l)}$ may thus have many different tuples of input  bits and output bits such that we may interpret $P^{(l)}$ as a probabilistic gate by proper global normalization.  For example gates which have doubly stochastic $P^{(l)}$ tensors may be run either backwards or forwards (a doubly stochastic matrix is a stochastic matrix whose rows and columns sum to $1$.)  Another example is given by a fan-out gate in our original energy function model
\begin{equation}
E_f(S_1,S_2,S_3)=\left\{ \begin{array}{ll}
0 & {\rm if}~S_1=S_2=S_3 \\
\Delta & {\rm otherwise} . 
 \end{array} \right. \label{eq:ef}\end{equation}  
In this case we may interpret the $M^{(l)}(S_1,S_2,S_3)$ arising from this energy function as having input $S_1$ and outputs $S_2$ and $S_3$, input $S_2$ and outputs $S_1$ and $S_3$, or input $S_3$ and outputs $S_2$ and $S_3$.  Contrary to this, suppose that the energy function is that of the one-to-two fanout described in Eq.~(\ref{eq:fan-out}),
\begin{equation}
E_F(S_1,S_2,S_3)= -\Delta S_1 (S_2+S_3).
\end{equation}
The $M^{(l)}(S_1,S_2,S_3)$ for this energy function has only one possible orientation between inputs and output: $S_1$ is an input and $S_2$ and $S_3$ are outputs.  This in general will lead to the following question: given the possible input and outputs tuples for the $M^{(l)}$s is it possible to assign these in such a way as to interpret the resulting total tensor as a probabilistic circuit.  We will now show that this problem is $\NP$-complete and is therefore (unless $\P=\NP$), in general, intractable.

To see that the problem is $\NP$-complete proceed as follows.  Assign to each bit of our physical system a boolean variable, $x_i$.  Assume that a given logical energy function $E_l$ is involved in only a constant number of bits.  Since $E_l$ is involved in a constant number of bits we can explicitly calculate all of the sets of inputs and output bits for which $M^{(l)}$ can be globally normalized so as to make it a probabilistic gate.  The variables $x_i$s are going to represent whether the corresponding bit is associated is an input or and output.  In particular, TRUE, will represent that the bit is an input and FALSE will represent that the bit is an output.  Thus from the sets of input and output bits for which one can consistently label inputs to $M^{(l)}$ we can construct a boolean function on the relevant bits such that this function is true if and only if a valid assignment of boolean variables respects a possible input and outputs.  Thus, for example, for $E_f$ above in Eq.~(\ref{eq:ef}), the boolean function would be
\begin{equation}
 [(x_1 \wedge \neg x_2 \wedge \neg x_3) \vee (\neg x_1 \wedge \neg x_2 \wedge  x_3)\vee(\neg x_1 \wedge x_2 \wedge \neg x_3)] \label{eq:bool}
\end{equation}
Constructing one such boolean function for each logic gate we can then construct a boolean function on all bits $x_i$ which is the conjunction (logical and) of all of the gate boolean functions.   Thus we see that by calculating for each $E_l$ the possible combinations of inputs and outputs that have an allowed interpretation in terms of a probabilistic circuit, we have a one-to-one mapping of the problem to the problem of deciding whether there is a satisfying assignment for the boolean function made up of the conjunction of logical terms with only a constant number of involved bits.  In order to demonstrate that the problem is \NP-complete we need now only show that the boolean function we can construct in this manner are general enough so as to yield a version of the satisfaction problem which is \NP-complete.  In particular it is clear that this problem is in $\NP$, since we can evaluate in polynomial time whether a given assignment of inputs and outputs can be interpreted as a probabilistic circuit.  To demonstrate $\NP$-completeness we need to show that it is at least as hard as another $\NP$-complete problem. 

We will show that the problem is $\NP$-complete by showing that some instance of the problem can be mapped to the one-in-three satisfiability problem of Schaefer~\cite{Schaefer:78a} (now sometimes called the monotone one-in-three satisfiability problem.)  Define $R(x,y,z)$ as the boolean function which is true iff exactly one of $(x,y,z)$ is true.  Notice that this is exactly the boolean formula of Eq.~(\ref{eq:bool}) (with $x_1=x$, $x_2=y$, and $x_3=z$.)  In the one-in-three satisfiability problem one is given a boolean function which is the conjunction (logical and) of clauses each of which are of the form $R(x,y,z)$ for a choice of variables $x,y,z$ in the problem.   Schaefer proved that the one-in-three satisfiability problem is \NP-complete.  Actually there is a slight subtlety here as the problem Schaefer considers must allow repeated variables. We will now show that it is possible to take a one-in-three satisfiability problem and convert it into an instance of the problem of spotting whether a given energy function can support interpretation as a classical probabilistic circuit. 

As mentioned above the one-in-three satisfiability problem has as input a conjunction of boolean functions which are all of the form $R(x,y,z)$, and where some boolean variables can be repeated,
\begin{equation}
f(x_1,\dots,x_n) = \bigwedge_{j=1}^r R(a_i,b_i,c_i), \label{eq:logicf}
\end{equation}
where $a_i,b_i,c_i \in \{x_1,\dots,x_n\}$ and none of the $a_i$, $b_i$, and $c_i$ are the same variable for fixed $i$.  Let $X_i=\{a_j | a_j=x_i\} \cup \{b_j |b_j=x_i\} \cup \{c_j|c_j=x_i\}$ denote the set of $a,b,c$ variables that are $x_i$.  We will define an energy function on $3r+\sum_{i=1}^n |X_i|$ spins which, when converted into the problem of trying to assign consistent inputs and outputs will yield exactly the boolean function $f$ of Eq.~(\ref{eq:logicf}).  Call the first $3r$ spins $S_i$ and the last $X_{tot}=\sum_{i=1}^n |X_i|$ spins $T_i$.   First define the energy function
\begin{equation}
E_M(S_1,\dots,S_{3r})= \sum_{j=1}^r E_f(S_{3j-2},S_{3j-1} ,S_{3j}),
\end{equation}
where $E_f$ is the fan-out energy function defined in Eq.~(\ref{eq:logicf}).  When one converts this into a logical expression for whether the spins are inputs or outputs, one gets exactly the one-in-three logical clause of Eq.~(\ref{eq:bool}).  

However, at this point, the variables that will arise in these clauses are all different.  This is what the other $X_{tot}$ spins are for.  Define the following energy function for the ``identity'' gate:
\begin{equation}
E_I^{(k)}(P_1,P_2,\dots,P_k,Q_1,Q_2,\dots,Q_k)= \left\{ \begin{array}{ll}
0 & {\rm if}~P_1=Q_1, P_2=Q_2,\dots, P_k=Q_k \\
\Delta & {\rm otherwise}  \end{array} \right. .
\end{equation}
This energy function corresponds to a logical reversible gate which acts as identity from the $P_i$ spins to the $Q_i$ spins.  This energy function has the property that the $P_i$ variables are all either inputs or all outputs (and the $Q_i$ variables are also all either inputs or outputs, opposite to the $P_i$ variable assignment.)  Thus we can use gates of this form to effectively make the spins of the $E_M$ gates constructed above the same variable.  In particular define the following energy function
\begin{equation}
E_e(S_1,\dots,S_{3r},T_1,\dots,T_{X_{tot}})=\sum_{i=1}^n E_I^{(|X_i|)}( S_{X_i[1]},S_{X_i[2]},\dots,S_{X_i[|X_i|]},T_{X_{t}^{(i)}},T_{X_{t}^{(i)}+1},\dots,T_{X_{t}^{(i)}+|X_i|}),
\end{equation}
where $X_{t}^{(i)}=\sum_{j<i} |X_i|$ and $X_i[j]$ is the $j$th element of $X_i$ translated over to a $S_{i}$ variable, i.e. $a_i$ corresponds to $S_{3i-2}$, $b_i$ corresponds to $S_{3i-1}$ and $c_i$ corresponds to $S_{3i}$.  

Now consider the energy function
\begin{equation}
E(S_1,\dots,S_{3r},T_1,\dots,T_{X_{tot}})= E_e(S_1,\dots,S_{3r},T_1,\dots,T_{X_{tot}}))+E_M(S_1\dots,S_{3r}),
\end{equation}
and stipulate that every term in the sum of $E_e$ will correspond to one gate and every term in $E_M$ will correspond to one gate.  Then if we convert this problem over into a boolean satisfaction problem as describe above, where each spin is a boolean variable representing whether the spin is an input or an output, we will obtain a boolean expression that is equivalent to the problem described in Eq.~(\ref{eq:logicf}).  Thus we have shown that we can, for a given one-in-three satisfaction problem, Eq.~(\ref{eq:logicf}) construct an energy function, along with labeling of the terms in this function corresponding to different gates, such that determining whether this energy function can be reinterpreted as a probabilistic computation is equivalent to the original one-in-three satisfaction problem.  This implies that the problem of determining whether such probabilistic computations exist is at least as hard as this $\NP$-complete problem.  Combined with the fact the problem is in $\NP$ this means that the problem of deciding whether a given energy function can be thought of as enacting a probabilistic circuit is $\NP$-complete.

Thus we have shown that the problem of determining a many-spin statistical mechanics system enacts a ground state spin computation, even given the decomposition of this system into logical gates, is \NP-complete.  We note in passing that this result is of perhaps some philosophical significance: determining whether a system can be thought of as a computer is shown to be computationally intractable (assuming $\P \neq \NP$).  

\subsection{Models that have a Probabilistic Gate Interpretation}

So far we have considered spin models in a very general setting unrelated to physical theories.  We point out, here, however, that there are a variety of models that have arisen in physics which admit such an interpretation.  We have already seen that that one dimensional Ising model admits an interpretation as a series of identity gates which err with some probability.  Here we point out some other models which admit such interpretations.  

\subsubsection{Ising Models on Trees}

A tree is a connected graph, $T=(V,E)$, with vertex set $V$, edge set $E$, and no cycles.  Assign to each spin a vertex, $S_v$ for $v \in V$.  To each edge $e=(v,w)$ assign an non-zero energy for the Ising coupling, $J : E \rightarrow \mathbb{R}-\{0\}$.  Then the Ising model on the tree has an energy function given by
\begin{equation}
E(s_1,\dots,s_{|V|})=\sum_{(v,w)\in E} J_{(v,w)} S_v S_w.
\end{equation}
Fix a root vertex $v_r \in V$.  First consider the model at zero temperature.  At $T=0$, following our discussion of the fan-out in Section~\ref{sec:general}, it is easy to see that the ground state of this model can be thought of as a series of fan-out gates originating from the root vertex $v_r$ with possible bit flips applied to the fan-out depending on the sign of $J_{(v,w)}$.  At $T \neq 0$ we see that, as in the discussion of the fan-out gate, each of the individual fan-out wires (with possible bit flips), will fail independently with a probability of
\begin{equation}
p_{(S_v,S_w)}={e^{\beta J_{(v,w)}}   \over e^{\beta J_{(v,w)}}+e^{-\beta J_{(v,w)}}}.
\end{equation}
Note that a failure for an antiferromagnetic coupling (bit flip) corresponds to an identity gate.  Ising models on trees have been studied in a variety of settings~\cite{Lyons:89a,Viteri:09a} and in fact the interpretation of this model in terms of a probabilistic circuit has been used extensively in the computer science literature, where the model goes under the name of {\em broadcasting}~\cite{Evans:00a}.

\subsubsection{$Z_2$ Lattice Pure Gauge Theories}

Perhaps slightly more interesting that Ising models on trees is the fact that $Z_2$ lattice gauge theories~\cite{Balian:75a,Cruetz:79a} can be cast as a probabilistic gate circuits.  For definiteness, first consider a (pure) $Z_2$ lattice gauge theory on a two dimensional square lattice with closed boundaries.  In this model one places spins on the edges of the lattice and defines the energy function
\begin{equation}
E=  \sum_{p \in P} \sum_{S_1,S_2,S_3,S_4 \in \delta p} J_p S_1 S_2 S_3 S_4 
\end{equation}
where $P$ is the set of all plaquettes of the square lattice, and $\delta p$ are the spins surrounding plaquette $p$.  Consider an individual term in this sum, and assume for simplicity that for all plaquettes, $p \in P$ that $J_p=J<0$.  We can view the individual term $J S_1 S_2 S_3 S_4$ as a gate element with either $0$, $1$, $2$, or $3$ inputs and $4$, $3$, $2$, and $1$ outputs respectively.  For example if we view it as a gate with $2$ inputs and $2$ outputs, then, in its ground state this corresponds to a probabilistic gate which, with equal probability, changes the input to a set of spins with the same parity of $-1$s as the input.  In other words the logical evolution is
\begin{eqnarray}
(+1,+1) &\rightarrow& 50\%~(+1,+1)~{\rm and~} 50\%~(-1,-1) \nonumber \\
(+1,-1) &\rightarrow& 50\%~(+1,-1)~{\rm and~} 50\%~(-1,+1) \nonumber \\
(-1,+1) &\rightarrow& 50\%~(-1,+1)~{\rm and~} 50\%~(-1,+1) \nonumber \\
(-1,-1) &\rightarrow& 50\%~(+1,+1)~{\rm and~} 50\%~(-1,-1).
\end{eqnarray}
The gate with $3$ inputs and $1$ outputs is deterministic and is given by $f(S_1,S_2,S_3)=S_1 S_2 S_3$.  The gate with $0$ inputs corresponds to a preparation of $4$ spins which have an even number of $-1$ spins and this state is prepared with equal probability.  The gate with $1$ input and $3$ outputs is a probabilistic gate which outputs an even number of $-1$s if the input is $+1$ all such possibilities occurring with equal probability.  Similarly it outputs an odd number of $-1$s if the input is $-1$ all such possibilities occurring with equal probability.  By picking boundary terms to serve as input and output bits, it is clear that it is possible to order the energy terms in the above sum in such a way that we can interpret the ground state as a probabilistic cellular automata like model using the above gates.  At finite gate these probabilistic gates all fail with some probability, where failure results in an equally likely failed output for the gate.  While we have define this for the two dimensional square lattice, it is clear that the above construction can result in a probabilistic circuit for a far greater number of lattices, the main criteria being that there exists a way to label inputs and outputs in such a way that a proper circuit is constructed.  

\section{Making Classical Ground State Spin Computation Robust at Finite Temperature} \label{sec:ft}

Having shown how the computed circuit energy functions give rise to thermal ensembles which can be interpreted as probabilistic circuits with probabilistic inputs, we now discuss how it is possible to make the model of classical ground state spin computation fault-tolerant at finite temperature.  Indeed this is now rather simple given that we have a mapping between the behavior of these systems at finite temperature and probabilistic circuits which have gates and preparations failing with a fixed probability.  We will need, however, to use the more general class of energy functions considered in Section~\ref{sec:general}. 

The general theory of computing in the presence of gates which fail goes back at least as far as the work of von Neumann~\cite{vonNeumann:56a}.  Von Neumann considered a model in which each logic gate failed with exactly the probability $\epsilon$.  In order to get this model to compute reliably, von Neumann used two major techniques.  The first techniques is that one needs to suitably encode information: in order to do this one  encodes a single bit across a bundle of wires in a logic circuit and interprets this as $0$ (or $1$) when a majority of the wires are in the $0$ (or $1$.)  Given such an encoding it is then easy to compute in a parallel manner such that the effect of gates failing is simply to decrease the proportion of $0$s (or $1$s) in an encoded $0$ (or $1$.)  However, the fact that the gates fail does cause the ratio of correct bits in an encoded bundle of bits to decrease.  In order to deal with this von Neumann used a second technique whereby faulty gates were used to perform error correction.  Actually this portion of von Neumann's analysis is not quite satisfactory, as it requires the use of a random permutation.  However Pippenger~\cite{Pippenger:85a} has shown that one can de-randomize this construction (one can obtain reasonable parameters for this method by using the expanders defined in \cite{Lubotzky:86a}).  The end result of this is that one can show that if one wishes to compute a logical circuit of size $n$ to an accuracy $1-\delta$ (that is, have the circuit fail with probability $\delta$) with gates that fail at $\epsilon<\epsilon_t$ for some fixed threshold $\epsilon_t$, one can do this using a circuit with the unreliable gates which is only a $O(\log^k {1 \over \delta})$ larger, where $k$ is a fixed constant.  This effectively means that one can create for all effective purposes reliable logic gates out unreliable logic gates with an overhead which is logarithmic in the inverse of the error desired.   

Having described von Neumann's construction one can now see how it is possible to make classical ground state spin computing fault-tolerant.  In particular we have shown above how the thermal ensemble for such systems gives rise to probabilistic gates.  For gates constructed using energy functions like Eq.~(\ref{eq:energy}) we have shown that the thermal ensemble arising for these functions is equivalent to a circuit in which gates fail with probability ${(2^k-1)e^{-b} \over 1+(2^k-1) e^{-b}}$ if the gate has $k$ output bits.  For $k=1$ this is $e^{-b} \over 1+ e^{-b}$.  We would like to use this for von Neumann's $\epsilon$ failure probability.  One complication arises, however, which is that in von Neumann's model the fan-out gates do not fail.  These gates are needed during the error correcting stage of his procedure.  We get around this by using fan-out gates constructed from energy functions like those in Section~\ref{sec:general}.  In particular define the fan-out gate via Eq.~(\ref{eq:fan-out}).  As shown in Section~\ref{sec:general} this gives rise in our probabilistic gate setting in which each fan-out resulting wire fails with probability $e^{-b} \over e^{-b}+e^{b}$.  These independent failures can then be reinterpreted as errors for the gates that follow them in von Neumann's error correcting construction.  One should note that there is a subtlety here in that if we desire to recover von Neuman's error model where the gates fail with exactly probability $\epsilon$ then a complicated calculation is needed in order to figure out $\epsilon$ in this model.  However, for large enough $\beta$ the effect will mostly be that the error probabilities add such that $\epsilon$ will be the sum of the fan-out error and the gate error.  

Putting this together we can conclude that there is a critical temperature below which one can use fault-tolerant constructions to design ground state spin energy functions which compute correctly with probability $1-\delta$ by using $O(\log^k{1 \over \delta})$ more interactions in the energy functions.  This means that while the general model of ground state spin computing is not tolerant to errors, for a cleverly designed setup one can overcome this difficulty and build robust devices, assuming that the temperature is low enough to keep the error probability below the error threshold.

\section{Relationship to Computational Complexity Theory} \label{sec:complexity}

Now we turn to a relationship between our model and computational complexity.  This connection is provided because the ground state spin computing model at zero temperature is intimately related to the Cook-Levin theorem from computational complexity.    

\subsection{The Cook-Levin Theorem}

The Cook-Levin theorem~\cite{Cook:71a,Trakhtenbrot:84a} is a fundamental result in the theory of computational complexity.  Here we briefly review this result.  For more details one is referred, for example, to the classic textbook~\cite{Sipser:05a}.  

Traditionally the theory of computational complexity is founded upon the computational model of a Turing machine~\cite{Turing:36a}.  The Church-Turing thesis roughly states that the Turing machine model adequately captures the notion of what can be performed by any physical instantiation of a computer and decades of experience have yet to challenge this assumption.   A deterministic Turing machine consists of the machine and a linear tape onto which elements from a finite alphabet, $\Gamma$, can be written into different cells on the tape.  The machine itself has an finite set of internal states, $Q$, a method for reading a symbol written in a cell on the tape, a method to write a symbol in a cell on the tape, and the ability to move either left or right one cell.   Formally the tape is assumed to be infinite or at least infinitely extendable.  Initially an input, consisting of a string of symbols from a finite alphabet $\Sigma \subseteq \Gamma$ is placed on the tape and the physical machine is placed on the first symbol of this string.  The Turing machine is assumed to start itself with its internal states in a special state called the start state, $q_0 \in Q$.  A deterministic Turing machine then proceeds to ``compute'' as follows.  At each time step the deterministic Turing machine examines the symbol of the cell it currently occupies and its internal state.  Then conditional on these two variables the Turing machine writes a symbol onto cell (possibly from a larger, but still finite, alphabet than the alphabet of the initial strings) it occupies, moves either left or right, and modifies its internal state.  These rules can be specified by a function, the transition function, $\delta: Q \times \Gamma \rightarrow Q \times \Gamma \times \{L,R\}$, where $L$ and $R$ denote moving left or right.   The deterministic Turing machine does this indefinitely until it reaches one of a set of its internal states known as halt states.  These halt states are labeled as either ``reject'' or ``accept.''  If we initially input the string $w$ and the deterministic Turing machine halts in accept state, then we say the deterministic Turing machine accepts $w$.  Similarly if on input $w$ the machine halts in a reject state, then we say that the machine rejects $w$.  Note that on some strings the Turing machine may neither accept or reject a string, but continue indefinitely.  The language of a Turing machine is the set of all strings taken from the input alphabet for which the Turing machine accepts for that string being input.  

Deterministic Turing machines capture many important features of real modern computers.  For example, the running time of most algorithms on a variety of hardware are closely related to the similar problems cast in terms of deterministic Turing machines.  An important class of languages are those for which a deterministic Turing machine accepts a string from the language in a time bounded by a constant times a polynomial in the size of the string.  Such languages are said to be in the complexity class $\P$.  

In contrast to a deterministic Turing machine, non-deterministic Turing machines are a model of computing which is not thought to properly encapsulate what can be efficiently computed on a modern computer.  A non-determinisitic Turing machine works in a manner similar to a deterministic Turing machine: it has an internal state, it can read and write symbols onto a tape, etc.  The difference is that at each step in the computation, a non-deterministic Turing machine doesn't have to chose one new state, new symbol and direction to move, but may branch to multiple such triples.  The transition function is now not $\delta:Q \times \Gamma \rightarrow Q \times \Gamma \times \{L,R\}$ but instead is of the form $\delta:Q \times \Gamma \rightarrow P(Q \times \Gamma \times \{L,R\})$ where $P$ denotes the power set.  Non-deterministic Turing machines thus explore multiple pathways at once.  Such a Turing machine halts when one of the states of any of the multiple branches it is exploring enters a halt state.  Again we can define whether the initial input is accepted or rejected depending on whether the state where the computation halts is a reject or accept state.   Given this we can define the language of the non-determinstic Turing machine as the set of strings in which the machine accepts the given input.  The time complexity of a non-deterministic Turing machine is the number of steps of executed by the Turing machine irrespective of the number of branches explored by the machine.  

Because a non-deterministic Turing machine is allowed to explore multiple pathways in its computation it is not clear that such machines can be reasonably mapped (in terms of say, time complexity of problems) to modern computers.  This question can be put on a more formal footing by defining the complexity class $\NP$.  $\NP$ is the class of languages which can be recognized by a non-deterministic Turing machine in a time bounded by a constant times a polynomial in the size of the string being decided.  The statement of whether $\NP$ is strictly larger than $\P$ is the famous $\P=\NP$ problem, one of the most important open problems in the theory of computational complexity.  That being said, most researchers do believe that $\P \neq \NP$ and that the model of a non-deterministic Turing machine is more powerful than computers we can physically construct.  

Given the above brief overview of the basic models of deterministic and non-deterministic Turing machines we can now introduce the Cook-Levin theorem.  To define the Cook-Levin problem we introduce the $\SAT$ problem.  
\begin{quote}
{\bf Boolean Satisfaction Problem} (\SAT): Given as input a description of a Boolean function from $n$ boolean inputs to one output, $f : \{0,1\}^n \rightarrow \{0,1\}$, decide whether there exists a set of inputs $x_1,x_2,\dots,x_n$, with each $x_i \in \{0,1\}$, such that $f(x_1,x_2,\dots,x_n)=1$.  
\end{quote} 
Cast in terms of languages, the $\SAT$  language is the set of strings describing a Boolean function such that the Boolean function has some input which evaluates to $1$.  The Cook-Levin theorem then states that the $\SAT$ is complete for $\NP$.  This means that every language in $\NP$ can be converted by a deterministic Turing machine in polynomial time to a $\SAT$ language and every $\SAT$ language can be converted into a language in $\NP$.  Thus, in a real sense, $\SAT$ captures what it means for a lanuage to be in $\NP$.  For example, if we could efficiently solve a $\SAT$ problem in polynomial time we would be able to solve every problem in $\NP$ in polynomial time and vice versa.  This is especially important considering that numerous problems beside $\SAT$ have been identified as $\NP$-complete.  This class of problems are essentially ``the most difficult'' in $\NP$ and are correspondingly the problems for which no known polynomial time algorithms for all instances of the problems in the class are thought to exist.  

We are most interested in the proof of the Cook-Levin theorem.  Before reviewing this it is useful to provide another equivalent characterization of $\NP$ which uses, instead of a non-deterministic Turing machine, a prover and a verifier.  The prover is assumed to have unlimited computational power while the verifier is limited to running a polynomial time algorithm on a deterministic Turing machine.  A language, $L$, is in $\NP$ if the prover can supply to a verifier a polynomial sized proof $p$ that a instance $x$ is in the language.  This proof must always be correct, i.e. the verifier must always be convinced, by using a Turing machine from $\P$, that the instance $x$ is the in the language.  Similarly the prover must never be able to convince the verifier incorrectly.  In other words if $x$ is not in $L$, then the verifier must always reject.  For example $\SAT$ is in $\NP$ by this definition since the prover can simply supply a satisfying input to the boolean function if the function can be satisfied, and the verifier can compute the boolean function on this input in polynomial time.  Similarly if the function does not admit a satisfying assignment the verifier can never be convinced that the instance is not in the language because no string the prover can give will lead to a satisfied boolean function.  Note that in the prover-verifier setting we do not require that the prover try to convince the verifier that an instance $x$ is not in $L$, only that if it is true that $x$ is not in $L$ that the prover cannot convince the verifier that $x$ is in $L$.

Given the prover/verifier definition of $\NP$ we can now discuss the Cook-Levin theorem.  The Cook-Levin theorem states that $\SAT$ is $\NP$-complete.  A problem, $P$, is $\NP$-complete if and only if every problem in $\NP$ can be transformed into $P$ by a deterministic Turing machine in polynomial time.  Thus $\SAT$, since it is $\NP$-complete, captures much of what is difficult about problems in all of $\NP$: if there were a polynomial time algorithm for $\SAT$ then there would be a polynomial time algorithm for every problem in $\NP$.

The proof that $\SAT$ is $\NP$-complete has two directions.  We have already stated the first direction, that $\SAT$ is a $\NP$ problem.  The meat of the proof is in the other direction, that every problem in $\NP$ can be converted into a $\SAT$ problem.  Here we do this in a slightly non-standard manner, using the prover-verifier definition of $\NP$.  In particular for any problem, $L$, in $\NP$ we know that there exists a verifier who takes as input the problem instance, $x$ and a certificate $p$, runs these on a deterministic Turing machine and verifies whether the instance is in the language.  Imagine now the time history of such a verification by a Turing machine.  Initially the Turing machine is at the beginning of the string, in its initial state, and the input string is on the tape.  At the next time step, the machine can change its state, move left, right or stay in place, and possibly change the symbol on the tape where the machine is located.  Notice, however that the contents of a cell on the tape, whether the machine is at that location, and the internal state of the machine if it is on that cell at the next time step is a function only of the contents of the cell and its neighbors at the prior step, whether the machine was on one of those cells, and its internal state if it was.  In other words the computation can be described {\em locally} in space and time.  Given this we can now describe how to convert a verifier, which is just a deterministic Turing machine and the strings representing the instance $x$ and the proof certificate $p$ into a problem in $\SAT$.

Suppose that the verifiers deterministic Turing machine runs in time $p(n)$ where $n=|x|+|p|$ is the size of the instance and the proof and $p$ is a polynomial function.  Define the following boolean variables, $L_{i,j,k}$, $H_{i,k}$, $S_{qk}$ to represent the history of the deterministic Turing machine in verifying the proof.  
\begin{table}
\begin{tabular}{|l|l|}
\hline
Variable & Represents \\ \hline
$L_{i,j,k}$ & True if tape cell $-p(n) \leq i \leq p(n)$ contains symbol $j$ at time $0 \leq k \leq p(n)$ \\ 
 $H_{ik}$ & True if the Turing Machine is at  tape cell $-p(n) \leq i \leq p(n)$ at time $0 \leq k \leq p(n)$ \\
 $S_{qk}$ & True if Turing machine is in state $q$ at time $0 \leq k \leq p(n)$ \\ \hline
\end{tabular}\caption{Variables in the Cook-Levin theorem.}
\end{table}
We think about the first two variables as being arranged in a large $O(p(n)) \times (p(n)+1)$ tableau onto which the computational history of the deterministic Turing machine is carried out.  We can now define a $\SAT$ problem over these variables which is satisfiable iff the deterministic Turing machine accepts the input string $x,p$.  This $\SAT$ formula is conjunction (logical and) of a clauses which enforce (a) that the initial contents of the tape are $x$, (b) that the machine is in the correct initial state, (c) the initial location of the head of the tape, (d) that there is only one symbol per tape cell, one state per time, one tape position per time, (e) that the tape cell is unchanged unless it is the one which can be written upon, (f) the machine properly updates according to the Turing machine transition function, and (g) the machine finishes in an accepting state.  For example condition (b) consists of clauses with a single variable $S_{q_0,0}$ where $q_0$ is the initial state of the Turing machine.  The most complicated clause in this construction is the one that encodes the local update rule for the machine, (f).  This can be done by adding the clauses
\begin{equation}
(H_{i,k} \wedge S_{q,k} \wedge L_{i,j,k}) \rightarrow (H_{i+d,k+1} \wedge S_{q^\prime,k+1} \wedge L_{i,j^\prime,k+1}) 
\end{equation}
for valid transitions of the Turing machine (that is for $q,q^\prime,j,j^\prime,d$ representing a valid change of state, symbol, and direction of movement for the read/write head of the Turing machine.)  Putting this together one thus sees that once can take the deterministic verifier along with the instance and $x$ and construct a polynomial sized boolean formula $f$ which is satisfied iff the verifier accepts the proof on the instance.  Note that we have not forced a constraint on the proof input to the circuit, but that the definition of $\NP$ in a prover-verifier setting leads to the satisfiability of the formula $f$ iff the instance $x$ is in the language.  

We can now see that connection between the Cook-Levin theorem and the classical ground state spin computing model.  In particular a central part of the proof of this theorem is that one can construct a satisfiability formula directly from the history of a computation.  In a similar manner classical ground state spin computing constructs an energy function whose ground state is the history of a computation.  The conversion of a logic circuit to such an energy function is essentially the arithmetization technique of computational complexity~\cite{Babai:91a,Shamir:92a}.  Note however that the important role here of the input and outputs to the constructed formula: in the case where the output is forced, we are led to $\NP$-complete problems, but when the input is forced then we are led to efficient constructions.  Motivated by this it is interesting to consider how the above constructions works but now at finite temperature.

\subsection{A Promise-MA Problem From Classical Ground State Spin Computing} \label{sec:ma}

We will now show that it is possible to use classical ground state spin computing to define, in analogy with $3-\SAT$ a {\bf Promise-}$\MA$-complete problem.  $\MA$ stands for Merlin-Arthur and is essentially a probabilistic generalization of the complexity class $\NP$~\cite{Babai:85a,Goldwasser:86a}.  It is defined as the follows:
\begin{quote}
{\bf Promise-MA}: A language $L=L_{yes} \cup L_{no} \subset \Sigma^*$, $L_{yes} \cap L_{no}=\emptyset$, is a promise problem in Merlin-Arthur ({\bf Promise-}\MA) if there is a probabilistic polynomial-time Turing machine
$V$ and a polynomial $p$ such that for all strings $x \in L$,
\begin{itemize}
  \item If $x \in L_{yes}$ then there exists a $w$ such that $|w|=poly(|x|)$ and $V(x,w)$ accepts with probability  at least $2/3$.
  \item If $x \in L_{no}$ then for all $w$ such that $|w|=poly(|x|)$ then $V(x,w)$ accepts with probability  at most $1/3$.
\end{itemize}
\end{quote}
Here $\Sigma$ is the alphabet over which the language is defined.  This version of the definition of $\MA$ is a promise problem, since $L_{yes} \cup L_{no}$ does not necessarily cover all possible elements of $\Sigma^*$.   The reason that this class is called Merlin-Arthur is because it can be defined in terms of an interactive proof system between a computationally unbounded prover, Merlin, and a more limited verifier, Arthur, who can perform feasible probabilistic computations (polynomial time probabilistic computations with bounded error, $\BPP$.)   Merlin is trying to convince Arthur that a certain instance $x$ is in a language $L_{yes}$.  He does this by communicating a polynomial size proof $w$ to Arthur.  Arthur can then take this proof and run a polynomial time probabilistic computation on this proof the result of which is: if $x \in L_{yes}$, then Arthur is, with high probability, convinced the $x$ is in $L_{yes}$, while if $x \in L_{no}$ no matter what proof Merlin supplies Arthur cannot be convinced that $x \in L_{yes}$ except with a small probability.

We define the following problem:
\begin{quote}
{\bf Boundary Expectation Value (BEV)}:  Let $k$ be a positive integer and $\beta$ a $k$ digit number.  Suppose we are given an energy function on $n$ bits, $E(s_1,\dots,s_n)$ which can be written as a sum of energy functions 
\begin{equation}
E(s)=\sum_l E_l(s)
\end{equation}
in such a way that this energy function can be interpreted as a circuit energy function (see Eq.~(\ref{eq:energy})) with $j>0$ input bits and $1$ output bit, and each term in $E_l$ is specified with $k$ bits of precision.  Consider the thermal ensemble generated by this energy function at inverse temperature $\beta$.  We promise that either there exists input bits $i_1,i_2,\dots,i_j$ and output bit $t$ such that for the reduced probability distribution on these bits
\begin{equation}
Pr(i_1,i_2,\dots,i_j,t=1) >{2 \over 3} \label{eq:cond1}
\end{equation}
or for any input bits $i_1,i_2,\dots,i_j$, 
\begin{equation}
Pr(i_1,i_2,\dots,i_j,t=1) <{1 \over 3} \label{eq:cond2}
\end{equation}
The problem {\bf BEV} is to determine whether the first of these conditions, Eq.~(\ref{eq:cond1}), holds.  In other words yes instances of this problem satisfy Eq.~(\ref{eq:cond1}) for some inputs $i_1,\dots,i_j$ and no instances satisfy Eq.~(\ref{eq:cond2}) for all $i_1,\dots,i_j$.
\end{quote} 
We claim that {\bf BEV} is {\bf Promise-}\MA-complete.  The proof of this is fairly straightforward given our prior discussion of the Cook-Levin theorem.  

First we claim that ${\bf BEV}$ is in {\bf Promise-}\MA.  To show this we must show that there exists an interactive proof of the form described above for the problem {\bf BEV}.  To see this note that a computationally unbounded Merlin can sample from the thermal ensemble for an instance of this problem and can check whether there exists an input $i_1,\dots,i_ j$ such that Eq.~(\ref{eq:cond1}) holds or whether for all inputs $i_1,\dots,i_j$, Eq.~(\ref{eq:cond2}) holds.  This will then serve as the proof, $w$, in the definition of \MA.  Given the description of $\beta$ and the circuit energy function, Arthur can construct a probabilistic Turing machine which implements the circuit related to this energy function using our construction of a probabilistic circuit from Section~\ref{sec:probtrans}.  Then Arthur can run this probabilistic Turing machine on the proof that Merlin supplies.  It is clear that if $x$ is a yes instance of {\bf BEV}, then the proof $w$ Merlin supplies will convince Arthur that he has a yes instance of this interactive protocol.  Further if $x$ is a no instance of {\bf BEV} it is also clear that no proof that Merlin supplies will convince Arthur that $x$ is a yes instance.  Thus it is clear that ${\bf BEV} $ is in {\bf Promise-}\MA.

The other direction of the proof requires that we show that every problem in {\bf Promise-}\MA ~can be converted into a {\bf BEV} problem.  To do this we follow the idea of the Cook-Levin theorem in that, for a \MA problem $x \in L_{yes}$, we convert the probabilistic Turing machine $V$ that verifies the proof $w$ of the \MA problem into a thermal circuit for which the associated {\bf BEV} problem has $(x,w)$ as the satisfying input.  Similarly, for an \MA problem $x \in L_{no}$, the thermal circuit corresponding to $V$ must correspond to a no instance of {\bf BEV}, since otherwise there would exist inputs $w=i_1,i_2,\dots, i_j$ which would cause $V(x,w)$ to accept with probability greater than $1/3$.  Clearly to do this we will make use of our probabilistic circuit representation result from Section~\ref{sec:probtrans}. The main tool that we need in such a construction is that we need to be able to construct circuit energy functions which perform probabilistic gates of a form we desire, for a fixed inverse temperature $\beta$.  To do this we can proceed as follows.  We can use our fault-tolerant constructions to produce circuits which operate with near deterministic behavior.  Then in order to augment a deterministic computation we need a source of randomness, indeed in most standard definitions of a probabilistic Turing machine one requires randomness which is close to uniformly random across the random bits being used (or in other words one requires bits which are nearly equal probability of being in $0$ or $1$.)  To obtain such randomness in our constructions, notice that if we do not force our inputs to be $0$ or $1$ then there is, in a thermal ensemble, an equal probability to be $0$ and $1$.  This allows one to create random bits in an circuit energy function.  However in order to get this to work with the deterministic part of our Turing machine, we must find a method to get this randomness into {\em encoded} bits in our circuits.  This can be accomplished by simply taking the equal probability $0$ and $1$ bits and running them through the circuit which performs error correction on the bundle corresponding to these bits.  Thus we see that we can construct at finite $\beta$ and circuit energy function which is near deterministic and which can also have input bits which are nearly uniformly random.  Thus we can proceed just as in the Cook-Levin theorem.  For a tableau describing the history of the probabilistic Turing machine, $V$, (when this probablistic Turing machine is simply a deterministic Turing machine aided by bits of randomness, we can construct an energy circuit function which implements the probabilistic Turing machine spatially across this tableau. 

Thus we have shown that {\bf BEV} is {\bf Promise-}\MA-complete.  While the proof of this result was rather straightforward and followed quickly from understanding the Cook-Levin theorem it is interesting to note that very few \MA-complete problems are known.  In fact the only other non-natural problem which is {\bf Promise-}\MA-complete to our knowledge is the stoquastic 6-\SAT problem~\cite{Bravyi:09a}.  The nearest comparable result is the classic result for the complexity of finding the ground state and computing the partition function for the Ising model of Barahona~\cite{Barahona:82a}.

\section{Quantum Models}

Finally we would like to briefly mention the quantum models which show some similarity with our classical model.  In particular these models were the inspiration for investigating this problem and thus this serves as a good location to introduce the open problems concerning these problem.

One of the quantum models relevant to this discussion are universal adiabatic quantum computing schemes.  In these models one shows how to adiabatically drag a many-body quantum system from one easily preparable ground state to the ground state of another many-body quantum system whose ground state is a superposition over the history of a quantum circuit.  Thus, similar to our models, the end result is a ground state which encodes a computation: but in this case the computation is a superposition over the history of the computation.  These models show that adiabatic quantum computing, which previously had been only used for optimization problems~\cite{Farhi:00a,Farhi:01a,Childs:02a} can also be used for universal quantum computation~\cite{Aharonov:04a,Mizel:06a,Kempe:06a,Oliveira:05a,Love:08a}.  However the final state of these models are systems with small energy gaps and thus the effects of working at non-zero temperature will have a profound effect on these models.  Indeed it is for exactly this reason that such universal adiabatic quantum computing schemes are not known to be fault-tolerant (see, however, \cite{Lidar:08a}.)  Similar reasoning follows for an earlier model of ground state quantum computing due to Mizel and co-workers~\cite{Mizel:04a,Mizel:02a}.

The problem of having a system whose ground state correctly computes but whose excited states do not is exactly the problem we have addressed in this paper for the classical ground state spin computing.  An interesting question then, when considering the quantum models, is whether there is a similar reinterpretation quantum thermal ensembles as spatially enacted quantum computations.  This is an important open problem which might conceivably lead to fault-tolerant methods for adiabatic quantum computing.

\section{Conclusion} \label{sec:conc}

We have introduce a set of energy functions on many bits (spins) which have the property that their ground state can be thought of as a spatially distributed deterministic computation.  At temperature greater than zero we have shown that the thermal ensemble arising from these models can be reinterpreted as a spatially distributed probabilistic computation.  Further, above zero temperature we see that the gates of a ground state spin computer can become unreliable and fail to execute the desired computation with high-fidelity.  However with the mapping of the thermal ensemble to probabilistic circuits we have shown how it is possible to make versions of the desired deterministic circuits which are fault-tolerant.  We have shown that determining whether a given classical energy function can be thought of as enacting a ground state computation is \NP-complete.  Finally we have shown that a problem concerning the thermal ensembles arising in our model give rise to a rare complete problem for the complexity class {\bf Promise-}\MA.  The models we have considered here are mostly devoid of connections to actual physical systems of interest, besides connections to quantum-dot cellular automata.  An interesting an important open question about these models is whether they arise in naturally occurring physical systems, or a suitably engineered system.  Another interesting question is the rate at which the ground state spin model thermalizes: a proof that the system reaches thermal equilibrium in a time polynomial in the size of the circuit being implement would be another step towards making this model more physically relevant.

\begin{acknowledgments}

D.B. and E.C. are supported by NSF grants 0621621, 0803478, 0829937, and 091640 and DARPA grant FA9550-09-1-0044.  K.R.B is supported by Georgia Tech.  We acknowledge useful conversations with David Meyer and Maxwell Pierce.

\end{acknowledgments}


\begin{thebibliography}{49}%
\makeatletter
\providecommand \@ifxundefined [1]{%
 \@ifx{#1\undefined}
}%
\providecommand \@ifnum [1]{%
 \ifnum #1\expandafter \@firstoftwo
 \else \expandafter \@secondoftwo
 \fi
}%
\providecommand \@ifx [1]{%
 \ifx #1\expandafter \@firstoftwo
 \else \expandafter \@secondoftwo
 \fi
}%
\providecommand \natexlab [1]{#1}%
\providecommand \enquote  [1]{``#1''}%
\providecommand \bibnamefont  [1]{#1}%
\providecommand \bibfnamefont [1]{#1}%
\providecommand \citenamefont [1]{#1}%
\providecommand \href@noop [0]{\@secondoftwo}%
\providecommand \href [0]{\begingroup \@sanitize@url \@href}%
\providecommand \@href[1]{\@@startlink{#1}\@@href}%
\providecommand \@@href[1]{\endgroup#1\@@endlink}%
\providecommand \@sanitize@url [0]{\catcode `\\12\catcode `\$12\catcode
  `\&12\catcode `\#12\catcode `\^12\catcode `\_12\catcode `\%12\relax}%
\providecommand \@@startlink[1]{}%
\providecommand \@@endlink[0]{}%
\providecommand \url  [0]{\begingroup\@sanitize@url \@url }%
\providecommand \@url [1]{\endgroup\@href {#1}{\urlprefix }}%
\providecommand \urlprefix  [0]{URL }%
\providecommand \Eprint [0]{\href }%
\@ifxundefined \urlstyle {%
  \providecommand \doi  [0]{\begingroup \@sanitize@url \@doi}%
  \providecommand \@doi [1]{\endgroup \@@startlink {\doibase
  #1}doi:\discretionary {}{}{}#1\@@endlink }%
}{%
  \providecommand \doi  [0]{doi:\discretionary{}{}{}\begingroup
  \urlstyle{rm}\Url }%
}%
\providecommand \doibase [0]{http://dx.doi.org/}%
\providecommand \Doi [0]{\begingroup \@sanitize@url \@Doi }%
\providecommand \@Doi  [1]{\endgroup\@@startlink{\doibase#1}\@@Doi}%
\providecommand \@@Doi [1]{#1\@@endlink}%
\providecommand \selectlanguage [0]{\@gobble}%
\providecommand \bibinfo  [0]{\@secondoftwo}%
\providecommand \bibfield  [0]{\@secondoftwo}%
\providecommand \translation [1]{[#1]}%
\providecommand \BibitemOpen [0]{}%
\providecommand \bibitemStop [0]{}%
\providecommand \bibitemNoStop [0]{.\EOS\space}%
\providecommand \EOS [0]{\spacefactor3000\relax}%
\providecommand \BibitemShut  [1]{\csname bibitem#1\endcsname}%
\bibitem [{\citenamefont {Feynman}(1960)}]{Feynman:60a}%
  \BibitemOpen
  \bibfield  {author} {\bibinfo {author} {\bibfnamefont {R.}~\bibnamefont
  {Feynman}},\ }\href@noop {} {\bibfield  {journal} {\bibinfo  {journal}
  {Engineering and Science (California Institute of Technology)},\ }\textbf
  {\bibinfo {volume} {23}},\ \bibinfo {pages} {22} (\bibinfo {year}
  {1960})}\BibitemShut {NoStop}%
\bibitem [{\citenamefont {Kish}(2002)}]{Kish:02a}%
  \BibitemOpen
  \bibfield  {author} {\bibinfo {author} {\bibfnamefont {L.~B.}\ \bibnamefont
  {Kish}},\ }\href@noop {} {\bibfield  {journal} {\bibinfo  {journal} {Phys.
  Lett. A},\ }\textbf {\bibinfo {volume} {305}},\ \bibinfo {pages} {144}
  (\bibinfo {year} {2002})}\BibitemShut {NoStop}%
\bibitem [{\citenamefont {Feynman}(1985)}]{Feynman:85a}%
  \BibitemOpen
  \bibfield  {author} {\bibinfo {author} {\bibfnamefont {R.}~\bibnamefont
  {Feynman}},\ }\href@noop {} {\bibfield  {journal} {\bibinfo  {journal}
  {Optics News},\ \bibinfo {pages} {11}} (\bibinfo {year} {1985})}\BibitemShut
  {NoStop}%
\bibitem [{\citenamefont {Benioff}(1982){\natexlab{a}}}]{Benioff:82a}%
  \BibitemOpen
  \bibfield  {author} {\bibinfo {author} {\bibfnamefont {P.}~\bibnamefont
  {Benioff}},\ }\href@noop {} {\bibfield  {journal} {\bibinfo  {journal} {J.
  Stat. Phys.},\ }\textbf {\bibinfo {volume} {29}},\ \bibinfo {pages} {515}
  (\bibinfo {year} {1982}{\natexlab{a}})}\BibitemShut {NoStop}%
\bibitem [{\citenamefont {Benioff}(1982){\natexlab{b}}}]{Benioff:82b}%
  \BibitemOpen
  \bibfield  {author} {\bibinfo {author} {\bibfnamefont {P.}~\bibnamefont
  {Benioff}},\ }\href@noop {} {\bibfield  {journal} {\bibinfo  {journal} {Phys.
  Rev. Lett.},\ }\textbf {\bibinfo {volume} {48}},\ \bibinfo {pages} {1581}
  (\bibinfo {year} {1982}{\natexlab{b}})}\BibitemShut {NoStop}%
\bibitem [{\citenamefont {Deutsch}(1985)}]{Deutsch:85a}%
  \BibitemOpen
  \bibfield  {author} {\bibinfo {author} {\bibfnamefont {D.}~\bibnamefont
  {Deutsch}},\ }\href@noop {} {\bibfield  {journal} {\bibinfo  {journal} {Proc.
  Roy. Soc. London Ser. A},\ }\textbf {\bibinfo {volume} {400}},\ \bibinfo
  {pages} {97} (\bibinfo {year} {1985})}\BibitemShut {NoStop}%
\bibitem [{\citenamefont {Shor}(1994)}]{Shor:94a}%
  \BibitemOpen
  \bibfield  {author} {\bibinfo {author} {\bibfnamefont {P.~W.}\ \bibnamefont
  {Shor}},\ }in\ \href@noop {} {\emph {\bibinfo {booktitle} {Proceedings of the
  35th Annual Symposium on the Foundations of Computer Science}}},\ \bibinfo
  {editor} {edited by\ \bibinfo {editor} {\bibfnamefont {S.}~\bibnamefont
  {Goldwasser}}}\ (\bibinfo  {publisher} {IEEE Computer Society},\ \bibinfo
  {address} {Los Alamitos, CA},\ \bibinfo {year} {1994})\ pp.\ \bibinfo {pages}
  {124--134}\BibitemShut {NoStop}%
\bibitem [{\citenamefont {von Neumann}(1956)}]{vonNeumann:56a}%
  \BibitemOpen
  \bibfield  {author} {\bibinfo {author} {\bibfnamefont {J.}~\bibnamefont {von
  Neumann}},\ }in\ \href@noop {} {\emph {\bibinfo {booktitle} {Automata
  Studies}}}\ (\bibinfo  {publisher} {Princeton University Press},\ \bibinfo
  {address} {Princeton, NJ},\ \bibinfo {year} {1956})\ pp.\ \bibinfo {pages}
  {329--378}\BibitemShut {NoStop}%
\bibitem [{\citenamefont {Lent}\ \emph {et~al.}(1993)\citenamefont {Lent},
  \citenamefont {Tougaw}, \citenamefont {Porod},\ and\ \citenamefont
  {Bernstein}}]{Lent:93a}%
  \BibitemOpen
  \bibfield  {author} {\bibinfo {author} {\bibfnamefont {C.}~\bibnamefont
  {Lent}}, \bibinfo {author} {\bibfnamefont {P.}~\bibnamefont {Tougaw}},
  \bibinfo {author} {\bibfnamefont {W.}~\bibnamefont {Porod}}, \ and\ \bibinfo
  {author} {\bibfnamefont {G.}~\bibnamefont {Bernstein}},\ }\href@noop {}
  {\bibfield  {journal} {\bibinfo  {journal} {Nanotechnology},\ }\textbf
  {\bibinfo {volume} {4}},\ \bibinfo {pages} {49} (\bibinfo {year}
  {1993})}\BibitemShut {NoStop}%
\bibitem [{\citenamefont {Lent}\ \emph {et~al.}(1994)\citenamefont {Lent},
  \citenamefont {Tougaw},\ and\ \citenamefont {Porod}}]{Lent:94a}%
  \BibitemOpen
  \bibfield  {author} {\bibinfo {author} {\bibfnamefont {C.}~\bibnamefont
  {Lent}}, \bibinfo {author} {\bibfnamefont {P.}~\bibnamefont {Tougaw}}, \ and\
  \bibinfo {author} {\bibfnamefont {W.}~\bibnamefont {Porod}},\ }in\ \href@noop
  {} {\emph {\bibinfo {booktitle} {Proceedings of 1994 Conference on Physics
  and Computation}}}\ (\bibinfo  {publisher} {IEEE Computer Society Press},\
  \bibinfo {year} {1994})\ pp.\ \bibinfo {pages} {5--13}\BibitemShut {NoStop}%
\bibitem [{\citenamefont {Wang}\ and\ \citenamefont
  {Lieberman}(2004)}]{Wang:04a}%
  \BibitemOpen
  \bibfield  {author} {\bibinfo {author} {\bibfnamefont {Y.}~\bibnamefont
  {Wang}}\ and\ \bibinfo {author} {\bibfnamefont {M.}~\bibnamefont
  {Lieberman}},\ }\href@noop {} {\bibfield  {journal} {\bibinfo  {journal}
  {IEEE Transactions on Nanotechnology},\ }\textbf {\bibinfo {volume} {3}},\
  \bibinfo {pages} {368} (\bibinfo {year} {2004})}\BibitemShut {NoStop}%
\bibitem [{\citenamefont {Evans}\ \emph {et~al.}(2000)\citenamefont {Evans},
  \citenamefont {Kenyon}, \citenamefont {Peres},\ and\ \citenamefont
  {Schulman}}]{Evans:00a}%
  \BibitemOpen
  \bibfield  {author} {\bibinfo {author} {\bibfnamefont {W.}~\bibnamefont
  {Evans}}, \bibinfo {author} {\bibfnamefont {C.}~\bibnamefont {Kenyon}},
  \bibinfo {author} {\bibfnamefont {Y.}~\bibnamefont {Peres}}, \ and\ \bibinfo
  {author} {\bibfnamefont {L.~J.}\ \bibnamefont {Schulman}},\ }\href@noop {}
  {\bibfield  {journal} {\bibinfo  {journal} {The Annals of Applied
  Probability},\ }\textbf {\bibinfo {volume} {10}},\ \bibinfo {pages} {410}
  (\bibinfo {year} {2000})},\ ISSN \bibinfo {issn} {10505164}\BibitemShut
  {NoStop}%
\bibitem [{\citenamefont {Mizel}\ \emph {et~al.}(2002)\citenamefont {Mizel},
  \citenamefont {Mitchell},\ and\ \citenamefont {Cohen}}]{Mizel:02a}%
  \BibitemOpen
  \bibfield  {author} {\bibinfo {author} {\bibfnamefont {A.}~\bibnamefont
  {Mizel}}, \bibinfo {author} {\bibfnamefont {M.~W.}\ \bibnamefont {Mitchell}},
  \ and\ \bibinfo {author} {\bibfnamefont {M.~L.}\ \bibnamefont {Cohen}},\
  }\href@noop {} {\bibfield  {journal} {\bibinfo  {journal} {Phys. Rev. A},\
  }\textbf {\bibinfo {volume} {65}},\ \bibinfo {pages} {022315} (\bibinfo
  {year} {2002})}\BibitemShut {NoStop}%
\bibitem [{\citenamefont {Mizel}(2004)}]{Mizel:04a}%
  \BibitemOpen
  \bibfield  {author} {\bibinfo {author} {\bibfnamefont {A.}~\bibnamefont
  {Mizel}},\ }\href@noop {} {\bibfield  {journal} {\bibinfo  {journal} {Phys.
  Rev. A},\ }\textbf {\bibinfo {volume} {70}},\ \bibinfo {pages} {012304}
  (\bibinfo {year} {2004})}\BibitemShut {NoStop}%
\bibitem [{\citenamefont {Aharonov}\ \emph {et~al.}(2004)\citenamefont
  {Aharonov}, \citenamefont {van Dam}, \citenamefont {Kempe}, \citenamefont
  {Landau}, \citenamefont {Lloyd},\ and\ \citenamefont {Regev}}]{Aharonov:04a}%
  \BibitemOpen
  \bibfield  {author} {\bibinfo {author} {\bibfnamefont {D.}~\bibnamefont
  {Aharonov}}, \bibinfo {author} {\bibfnamefont {W.}~\bibnamefont {van Dam}},
  \bibinfo {author} {\bibfnamefont {J.}~\bibnamefont {Kempe}}, \bibinfo
  {author} {\bibfnamefont {Z.}~\bibnamefont {Landau}}, \bibinfo {author}
  {\bibfnamefont {S.}~\bibnamefont {Lloyd}}, \ and\ \bibinfo {author}
  {\bibfnamefont {O.}~\bibnamefont {Regev}},\ }in\ \href@noop {} {\emph
  {\bibinfo {booktitle} {45th Annual IEEE Symposium on Foundations of Computer
  Science}}}\ (\bibinfo  {publisher} {IEEE Computer Society},\ \bibinfo
  {address} {Los Alamitos, CA, USA},\ \bibinfo {year} {2004})\ pp.\ \bibinfo
  {pages} {42--51}\BibitemShut {NoStop}%
\bibitem [{\citenamefont {Mizel}\ \emph {et~al.}(2007)\citenamefont {Mizel},
  \citenamefont {Lidar},\ and\ \citenamefont {Mitchell}}]{Mizel:06a}%
  \BibitemOpen
  \bibfield  {author} {\bibinfo {author} {\bibfnamefont {A.}~\bibnamefont
  {Mizel}}, \bibinfo {author} {\bibfnamefont {D.~A.}\ \bibnamefont {Lidar}}, \
  and\ \bibinfo {author} {\bibfnamefont {M.~W.}\ \bibnamefont {Mitchell}},\
  }\href@noop {} {\bibfield  {journal} {\bibinfo  {journal} {Phys. Rev.
  Lett.},\ \bibinfo {pages} {070502}} (\bibinfo {year} {2007})},\ \Eprint
  {http://arxiv.org/abs/arXiv:quant-ph/0609067} {arXiv:quant-ph/0609067}
  \BibitemShut {NoStop}%
\bibitem [{\citenamefont {Kempe}\ \emph {et~al.}(2006)\citenamefont {Kempe},
  \citenamefont {Kitaev},\ and\ \citenamefont {Regev}}]{Kempe:06a}%
  \BibitemOpen
  \bibfield  {author} {\bibinfo {author} {\bibfnamefont {J.}~\bibnamefont
  {Kempe}}, \bibinfo {author} {\bibfnamefont {A.}~\bibnamefont {Kitaev}}, \
  and\ \bibinfo {author} {\bibfnamefont {O.}~\bibnamefont {Regev}},\
  }\href@noop {} {\bibfield  {journal} {\bibinfo  {journal} {SIAM Journal of
  Computing},\ }\textbf {\bibinfo {volume} {35}},\ \bibinfo {pages} {1070}
  (\bibinfo {year} {2006})}\BibitemShut {NoStop}%
\bibitem [{\citenamefont {Oliveira}\ and\ \citenamefont
  {Terhal}(2008)}]{Oliveira:05a}%
  \BibitemOpen
  \bibfield  {author} {\bibinfo {author} {\bibfnamefont {R.}~\bibnamefont
  {Oliveira}}\ and\ \bibinfo {author} {\bibfnamefont {B.}~\bibnamefont
  {Terhal}},\ }\href@noop {} {\bibfield  {journal} {\bibinfo  {journal}
  {Quantum Inform. Compu.},\ }\textbf {\bibinfo {volume} {8}},\ \bibinfo
  {pages} {0900} (\bibinfo {year} {2008})}\BibitemShut {NoStop}%
\bibitem [{\citenamefont {Biamonte}\ and\ \citenamefont
  {Love}(2008)}]{Love:08a}%
  \BibitemOpen
  \bibfield  {author} {\bibinfo {author} {\bibfnamefont {J.~D.}\ \bibnamefont
  {Biamonte}}\ and\ \bibinfo {author} {\bibfnamefont {P.~J.}\ \bibnamefont
  {Love}},\ }\href@noop {} {\bibfield  {journal} {\bibinfo  {journal} {Phys.
  Rev. A},\ }\textbf {\bibinfo {volume} {78}},\ \bibinfo {pages} {012352}
  (\bibinfo {year} {2008})}\BibitemShut {NoStop}%
\bibitem [{\citenamefont {Babai}(1985)}]{Babai:85a}%
  \BibitemOpen
  \bibfield  {author} {\bibinfo {author} {\bibfnamefont {L.}~\bibnamefont
  {Babai}},\ }in\ \href@noop {} {\emph {\bibinfo {booktitle} {Proceedings of
  the 17th Annual ACM Symposium on Theory of Computing}}}\ (\bibinfo
  {publisher} {ACM Press},\ \bibinfo {year} {1985})\ pp.\ \bibinfo {pages}
  {421--429}\BibitemShut {NoStop}%
\bibitem [{\citenamefont {Goldwasser}\ and\ \citenamefont
  {Sipser}(1986)}]{Goldwasser:86a}%
  \BibitemOpen
  \bibfield  {author} {\bibinfo {author} {\bibfnamefont {S.}~\bibnamefont
  {Goldwasser}}\ and\ \bibinfo {author} {\bibfnamefont {M.}~\bibnamefont
  {Sipser}},\ }in\ \href@noop {} {\emph {\bibinfo {booktitle} {Proceedings of
  the 18th Annual ACM Symposium on Theory of Computing}}}\ (\bibinfo
  {publisher} {ACM Press},\ \bibinfo {year} {1986})\ pp.\ \bibinfo {pages}
  {59--68}\BibitemShut {NoStop}%
\bibitem [{\citenamefont {Cook}(1971)}]{Cook:71a}%
  \BibitemOpen
  \bibfield  {author} {\bibinfo {author} {\bibfnamefont {S.}~\bibnamefont
  {Cook}},\ }in\ \href@noop {} {\emph {\bibinfo {booktitle} {Proceedings of the
  3rd Annual ACM Symposium on Theory of Computing}}}\ (\bibinfo {year} {1971})\
  p.\ \bibinfo {pages} {151}\BibitemShut {NoStop}%
\bibitem [{\citenamefont {Trakhtenbrot}(1984)}]{Trakhtenbrot:84a}%
  \BibitemOpen
  \bibfield  {author} {\bibinfo {author} {\bibfnamefont {B.~A.}\ \bibnamefont
  {Trakhtenbrot}},\ }\href@noop {} {\bibfield  {journal} {\bibinfo  {journal}
  {IEEE Annals of the History of Computing},\ }\textbf {\bibinfo {volume}
  {6}},\ \bibinfo {pages} {384} (\bibinfo {year} {1984})}\BibitemShut {NoStop}%
\bibitem [{\citenamefont {Bacon}\ and\ \citenamefont
  {Flammia}(2009)}]{Bacon:09a}%
  \BibitemOpen
  \bibfield  {author} {\bibinfo {author} {\bibfnamefont {D.}~\bibnamefont
  {Bacon}}\ and\ \bibinfo {author} {\bibfnamefont {S.~T.}\ \bibnamefont
  {Flammia}},\ }\href@noop {} {\bibfield  {journal} {\bibinfo  {journal} {Phys.
  Rev. Lett.},\ }\textbf {\bibinfo {volume} {103}},\ \bibinfo {pages} {120504}
  (\bibinfo {year} {2009})}\BibitemShut {NoStop}%
\bibitem [{\citenamefont {Bacon}\ and\ \citenamefont
  {Flammia}(2010)}]{Bacon:09b}%
  \BibitemOpen
  \bibfield  {author} {\bibinfo {author} {\bibfnamefont {D.}~\bibnamefont
  {Bacon}}\ and\ \bibinfo {author} {\bibfnamefont {S.~T.}\ \bibnamefont
  {Flammia}},\ }\href@noop {} {\enquote {\bibinfo {title} {Adiabatic cluster
  state quantum computing,},}\ }\bibinfo {howpublished} {arXiv:0912.2098}
  (\bibinfo {year} {2010})\BibitemShut {NoStop}%
\bibitem [{\citenamefont {Oreshkov}\ \emph {et~al.}(2009)\citenamefont
  {Oreshkov}, \citenamefont {Brun},\ and\ \citenamefont
  {Lidar}}]{Oreshkov:09a}%
  \BibitemOpen
  \bibfield  {author} {\bibinfo {author} {\bibfnamefont {O.}~\bibnamefont
  {Oreshkov}}, \bibinfo {author} {\bibfnamefont {T.~A.}\ \bibnamefont {Brun}},
  \ and\ \bibinfo {author} {\bibfnamefont {D.~A.}\ \bibnamefont {Lidar}},\
  }\href@noop {} {\bibfield  {journal} {\bibinfo  {journal} {Phys. Rev.
  Lett.},\ }\textbf {\bibinfo {volume} {102}},\ \bibinfo {pages} {070502}
  (\bibinfo {year} {2009})}\BibitemShut {NoStop}%
\bibitem [{\citenamefont {Oreshkov}(2009)}]{Oreshkov:09b}%
  \BibitemOpen
  \bibfield  {author} {\bibinfo {author} {\bibfnamefont {O.}~\bibnamefont
  {Oreshkov}},\ }\Doi {10.1103/PhysRevLett.103.090502} {\bibfield  {journal}
  {\bibinfo  {journal} {Phys. Rev. Lett.},\ }\textbf {\bibinfo {volume}
  {103}},\ \bibinfo {eid} {090502} (\bibinfo {year} {2009})}\BibitemShut
  {NoStop}%
\bibitem [{\citenamefont {Ising}(1925)}]{Ising:25a}%
  \BibitemOpen
  \bibfield  {author} {\bibinfo {author} {\bibfnamefont {E.}~\bibnamefont
  {Ising}},\ }\href@noop {} {\bibfield  {journal} {\bibinfo  {journal} {Z.
  Physik},\ }\textbf {\bibinfo {volume} {31}},\ \bibinfo {pages} {253}
  (\bibinfo {year} {1925})}\BibitemShut {NoStop}%
\bibitem [{\citenamefont {Ungarelli}\ \emph {et~al.}(2000)\citenamefont
  {Ungarelli}, \citenamefont {Francaviglia}, \citenamefont {Macucci},\ and\
  \citenamefont {Iannaccone}}]{Ungarelli:00a}%
  \BibitemOpen
  \bibfield  {author} {\bibinfo {author} {\bibfnamefont {C.}~\bibnamefont
  {Ungarelli}}, \bibinfo {author} {\bibfnamefont {S.}~\bibnamefont
  {Francaviglia}}, \bibinfo {author} {\bibfnamefont {M.}~\bibnamefont
  {Macucci}}, \ and\ \bibinfo {author} {\bibfnamefont {G.}~\bibnamefont
  {Iannaccone}},\ }\Doi {10.1063/1.372987} {\bibfield  {journal} {\bibinfo
  {journal} {Journal of Applied Physics},\ }\textbf {\bibinfo {volume} {87}},\
  \bibinfo {pages} {7320} (\bibinfo {year} {2000})}\BibitemShut {NoStop}%
\bibitem [{\citenamefont {Eggarter}(1974)}]{Eggarter:74a}%
  \BibitemOpen
  \bibfield  {author} {\bibinfo {author} {\bibfnamefont {T.~P.}\ \bibnamefont
  {Eggarter}},\ }\Doi {10.1103/PhysRevB.9.2989} {\bibfield  {journal} {\bibinfo
   {journal} {Phys. Rev. B},\ }\textbf {\bibinfo {volume} {9}},\ \bibinfo
  {pages} {2989} (\bibinfo {year} {1974})}\BibitemShut {NoStop}%
\bibitem [{\citenamefont {Markov}\ and\ \citenamefont
  {Shi}(2008)}]{Markov:08a}%
  \BibitemOpen
  \bibfield  {author} {\bibinfo {author} {\bibfnamefont {I.~L.}\ \bibnamefont
  {Markov}}\ and\ \bibinfo {author} {\bibfnamefont {Y.}~\bibnamefont {Shi}},\
  }\Doi {10.1137/050644756} {\bibfield  {journal} {\bibinfo  {journal} {SIAM
  Journal on Computing},\ }\textbf {\bibinfo {volume} {38}},\ \bibinfo {pages}
  {963} (\bibinfo {year} {2008})}\BibitemShut {NoStop}%
\bibitem [{\citenamefont {Wu}\ and\ \citenamefont {Sprung}(84)}]{Wu:98a}%
  \BibitemOpen
  \bibfield  {author} {\bibinfo {author} {\bibfnamefont {H.}~\bibnamefont
  {Wu}}\ and\ \bibinfo {author} {\bibfnamefont {D.}~\bibnamefont {Sprung}},\
  }\href@noop {} {\bibfield  {journal} {\bibinfo  {journal} {J. Appl. Phys},\
  \bibinfo {pages} {4000}} (\bibinfo {year} {84})}\BibitemShut {NoStop}%
\bibitem [{\citenamefont {Schaefer}(1978)}]{Schaefer:78a}%
  \BibitemOpen
  \bibfield  {author} {\bibinfo {author} {\bibfnamefont {T.~J.}\ \bibnamefont
  {Schaefer}},\ }in\ \Doi {http://doi.acm.org/10.1145/800133.804350} {\emph
  {\bibinfo {booktitle} {Proceedings of the tenth annual ACM symposium on
  Theory of computing}}}\ (\bibinfo  {publisher} {ACM},\ \bibinfo {address}
  {New York, NY, USA},\ \bibinfo {year} {1978})\ pp.\ \bibinfo {pages}
  {216--226}\BibitemShut {NoStop}%
\bibitem [{\citenamefont {Lyons}(1989)}]{Lyons:89a}%
  \BibitemOpen
  \bibfield  {author} {\bibinfo {author} {\bibfnamefont {R.}~\bibnamefont
  {Lyons}},\ }\href@noop {} {\bibfield  {journal} {\bibinfo  {journal} {Comm.
  Math. Phys.},\ }\textbf {\bibinfo {volume} {125}},\ \bibinfo {pages} {337}
  (\bibinfo {year} {1989})}\BibitemShut {NoStop}%
\bibitem [{\citenamefont {Viteri}\ \emph {et~al.}(2009)\citenamefont {Viteri},
  \citenamefont {Tomita},\ and\ \citenamefont {Brown}}]{Viteri:09a}%
  \BibitemOpen
  \bibfield  {author} {\bibinfo {author} {\bibfnamefont {C.~R.}\ \bibnamefont
  {Viteri}}, \bibinfo {author} {\bibfnamefont {Y.}~\bibnamefont {Tomita}}, \
  and\ \bibinfo {author} {\bibfnamefont {K.~R.}\ \bibnamefont {Brown}},\
  }\href@noop {} {\bibfield  {journal} {\bibinfo  {journal} {Phys. Rev. A},\
  }\textbf {\bibinfo {volume} {80}},\ \bibinfo {pages} {042313} (\bibinfo
  {year} {2009})}\BibitemShut {NoStop}%
\bibitem [{\citenamefont {Balian}\ \emph {et~al.}(1975)\citenamefont {Balian},
  \citenamefont {Drouffe},\ and\ \citenamefont {Itzykson}}]{Balian:75a}%
  \BibitemOpen
  \bibfield  {author} {\bibinfo {author} {\bibfnamefont {R.}~\bibnamefont
  {Balian}}, \bibinfo {author} {\bibfnamefont {J.~M.}\ \bibnamefont {Drouffe}},
  \ and\ \bibinfo {author} {\bibfnamefont {C.}~\bibnamefont {Itzykson}},\ }\Doi
  {10.1103/PhysRevD.11.2098} {\bibfield  {journal} {\bibinfo  {journal} {Phys.
  Rev. D},\ }\textbf {\bibinfo {volume} {11}},\ \bibinfo {pages} {2098}
  (\bibinfo {year} {1975})}\BibitemShut {NoStop}%
\bibitem [{\citenamefont {Creutz}\ \emph {et~al.}(1979)\citenamefont {Creutz},
  \citenamefont {Jacobs},\ and\ \citenamefont {Rebbi}}]{Cruetz:79a}%
  \BibitemOpen
  \bibfield  {author} {\bibinfo {author} {\bibfnamefont {M.}~\bibnamefont
  {Creutz}}, \bibinfo {author} {\bibfnamefont {L.}~\bibnamefont {Jacobs}}, \
  and\ \bibinfo {author} {\bibfnamefont {C.}~\bibnamefont {Rebbi}},\ }\Doi
  {10.1103/PhysRevLett.42.1390} {\bibfield  {journal} {\bibinfo  {journal}
  {Phys. Rev. Lett.},\ }\textbf {\bibinfo {volume} {42}},\ \bibinfo {pages}
  {1390} (\bibinfo {year} {1979})}\BibitemShut {NoStop}%
\bibitem [{\citenamefont {Pippenger}(1985)}]{Pippenger:85a}%
  \BibitemOpen
  \bibfield  {author} {\bibinfo {author} {\bibfnamefont {N.}~\bibnamefont
  {Pippenger}},\ }in\ \href@noop {} {\emph {\bibinfo {booktitle} {Proc. of the
  26th Annual Symposium on Foundations of Computer Science}}}\ (\bibinfo {year}
  {1985})\ pp.\ \bibinfo {pages} {30--38}\BibitemShut {NoStop}%
\bibitem [{\citenamefont {Lubotzky}\ \emph {et~al.}(1986)\citenamefont
  {Lubotzky}, \citenamefont {Phillips},\ and\ \citenamefont
  {Sarnak}}]{Lubotzky:86a}%
  \BibitemOpen
  \bibfield  {author} {\bibinfo {author} {\bibfnamefont {A.}~\bibnamefont
  {Lubotzky}}, \bibinfo {author} {\bibfnamefont {R.}~\bibnamefont {Phillips}},
  \ and\ \bibinfo {author} {\bibfnamefont {P.}~\bibnamefont {Sarnak}},\ }in\
  \Doi {http://doi.acm.org/10.1145/12130.12154} {\emph {\bibinfo {booktitle}
  {Proceedings of the eighteenth annual ACM symposium on Theory of
  computing}}}\ (\bibinfo  {publisher} {ACM},\ \bibinfo {address} {New York,
  NY, USA},\ \bibinfo {year} {1986})\ pp.\ \bibinfo {pages} {240--246},\ ISBN
  \bibinfo {isbn} {0-89791-193-8}\BibitemShut {NoStop}%
\bibitem [{\citenamefont {Sipser}(2005)}]{Sipser:05a}%
  \BibitemOpen
  \bibfield  {author} {\bibinfo {author} {\bibfnamefont {M.}~\bibnamefont
  {Sipser}},\ }\href@noop {} {\emph {\bibinfo {title} {Introduction to the
  Theory of Computation}}},\ \bibinfo {edition} {2nd}\ ed.\ (\bibinfo
  {publisher} {Course Technology},\ \bibinfo {year} {2005})\BibitemShut
  {NoStop}%
\bibitem [{\citenamefont {Turing}(1936)}]{Turing:36a}%
  \BibitemOpen
  \bibfield  {author} {\bibinfo {author} {\bibfnamefont {A.~M.}\ \bibnamefont
  {Turing}},\ }\href@noop {} {\bibfield  {journal} {\bibinfo  {journal} {Proc.
  Lond. Math. Soc. 2},\ }\textbf {\bibinfo {volume} {42}},\ \bibinfo {pages}
  {230} (\bibinfo {year} {1936})}\BibitemShut {NoStop}%
\bibitem [{\citenamefont {Babai}\ and\ \citenamefont
  {Fortnow}(1991)}]{Babai:91a}%
  \BibitemOpen
  \bibfield  {author} {\bibinfo {author} {\bibfnamefont {L.}~\bibnamefont
  {Babai}}\ and\ \bibinfo {author} {\bibfnamefont {L.}~\bibnamefont
  {Fortnow}},\ }\href@noop {} {\bibfield  {journal} {\bibinfo  {journal}
  {Compuational Complexity},\ }\textbf {\bibinfo {volume} {1}},\ \bibinfo
  {pages} {41} (\bibinfo {year} {1991})}\BibitemShut {NoStop}%
\bibitem [{\citenamefont {Shamir}(1992)}]{Shamir:92a}%
  \BibitemOpen
  \bibfield  {author} {\bibinfo {author} {\bibfnamefont {A.}~\bibnamefont
  {Shamir}},\ }\Doi {http://doi.acm.org/10.1145/146585.146609} {\bibfield
  {journal} {\bibinfo  {journal} {J. ACM},\ }\textbf {\bibinfo {volume} {39}},\
  \bibinfo {pages} {869} (\bibinfo {year} {1992})},\ ISSN \bibinfo {issn}
  {0004-5411}\BibitemShut {NoStop}%
\bibitem [{\citenamefont {Bravyi}\ and\ \citenamefont
  {Terhal}(2009)}]{Bravyi:09a}%
  \BibitemOpen
  \bibfield  {author} {\bibinfo {author} {\bibfnamefont {S.}~\bibnamefont
  {Bravyi}}\ and\ \bibinfo {author} {\bibfnamefont {B.}~\bibnamefont
  {Terhal}},\ }\Doi {10.1137/08072689X} {\bibfield  {journal} {\bibinfo
  {journal} {SIAM Journal on Computing},\ }\textbf {\bibinfo {volume} {39}},\
  \bibinfo {pages} {1462} (\bibinfo {year} {2009})}\BibitemShut {NoStop}%
\bibitem [{\citenamefont {Barahona}(1982)}]{Barahona:82a}%
  \BibitemOpen
  \bibfield  {author} {\bibinfo {author} {\bibfnamefont {F.}~\bibnamefont
  {Barahona}},\ }\href@noop {} {\bibfield  {journal} {\bibinfo  {journal} {J.
  Phys. A: Math. Gen.},\ }\textbf {\bibinfo {volume} {15}},\ \bibinfo {pages}
  {3241} (\bibinfo {year} {1982})}\BibitemShut {NoStop}%
\bibitem [{\citenamefont {Farhi}\ \emph {et~al.}(2000)\citenamefont {Farhi},
  \citenamefont {Goldstone}, \citenamefont {Gutmann},\ and\ \citenamefont
  {Sipser}}]{Farhi:00a}%
  \BibitemOpen
  \bibfield  {author} {\bibinfo {author} {\bibfnamefont {E.}~\bibnamefont
  {Farhi}}, \bibinfo {author} {\bibfnamefont {J.}~\bibnamefont {Goldstone}},
  \bibinfo {author} {\bibfnamefont {S.}~\bibnamefont {Gutmann}}, \ and\
  \bibinfo {author} {\bibfnamefont {M.}~\bibnamefont {Sipser}},\ }\href@noop {}
  {\enquote {\bibinfo {title} {Quantum computation by adiabatic evolution},}\ }
  (\bibinfo {year} {2000}),\ \Eprint
  {http://arxiv.org/abs/arXiv:quant-ph/0001106} {arXiv:quant-ph/0001106}
  \BibitemShut {NoStop}%
\bibitem [{\citenamefont {Farhi}\ \emph {et~al.}(2001)\citenamefont {Farhi},
  \citenamefont {Goldstone}, \citenamefont {Gutmann}, \citenamefont {Lapan},
  \citenamefont {Lundgren},\ and\ \citenamefont {Preda}}]{Farhi:01a}%
  \BibitemOpen
  \bibfield  {author} {\bibinfo {author} {\bibfnamefont {E.}~\bibnamefont
  {Farhi}}, \bibinfo {author} {\bibfnamefont {J.}~\bibnamefont {Goldstone}},
  \bibinfo {author} {\bibfnamefont {S.}~\bibnamefont {Gutmann}}, \bibinfo
  {author} {\bibfnamefont {J.}~\bibnamefont {Lapan}}, \bibinfo {author}
  {\bibfnamefont {A.}~\bibnamefont {Lundgren}}, \ and\ \bibinfo {author}
  {\bibfnamefont {D.}~\bibnamefont {Preda}},\ }\href@noop {} {\bibfield
  {journal} {\bibinfo  {journal} {Science},\ }\textbf {\bibinfo {volume}
  {292}},\ \bibinfo {pages} {472} (\bibinfo {year} {2001})}\BibitemShut
  {NoStop}%
\bibitem [{\citenamefont {Childs}\ \emph {et~al.}(2002)\citenamefont {Childs},
  \citenamefont {Farhi}, \citenamefont {Goldstone},\ and\ \citenamefont
  {Gutmann}}]{Childs:02a}%
  \BibitemOpen
  \bibfield  {author} {\bibinfo {author} {\bibfnamefont {A.}~\bibnamefont
  {Childs}}, \bibinfo {author} {\bibfnamefont {E.}~\bibnamefont {Farhi}},
  \bibinfo {author} {\bibfnamefont {J.}~\bibnamefont {Goldstone}}, \ and\
  \bibinfo {author} {\bibfnamefont {S.}~\bibnamefont {Gutmann}},\ }\href@noop
  {} {\bibfield  {journal} {\bibinfo  {journal} {Quantum Inform. Compu.},\
  }\textbf {\bibinfo {volume} {2}},\ \bibinfo {pages} {181} (\bibinfo {year}
  {2002})}\BibitemShut {NoStop}%
\bibitem [{\citenamefont {Lidar}(2008)}]{Lidar:08a}%
  \BibitemOpen
  \bibfield  {author} {\bibinfo {author} {\bibfnamefont {D.~A.}\ \bibnamefont
  {Lidar}},\ }\href@noop {} {\bibfield  {journal} {\bibinfo  {journal} {Phys.
  Rev. Lett.},\ }\textbf {\bibinfo {volume} {100}},\ \bibinfo {pages} {160506}
  (\bibinfo {year} {2008})}\BibitemShut {NoStop}%
\end{thebibliography}
 \end{document}